\documentclass[fleqn,useAMS,usenatbib]{mnras}

\usepackage{amsmath}
\usepackage{amssymb}
\usepackage{graphicx}
\usepackage{times}
\usepackage{hyperref}

\title[SMBH mergers in FLARES]
{First Light And Reionization Epoch Simulations (FLARES) -- XIX: Supermassive black hole mergers in the early Universe and their environmental dependence}

\author[S. Liao et al.]
{Shihong Liao,$^{1}$\thanks{Email: shliao@nao.cas.cn} Dimitrios Irodotou,$^{2}$ Maxwell G. A. Maltz,$^{3}$ Christopher C. Lovell,$^{4}$ \newauthor Zhen Jiang,$^{1}$ Sophie L. Newman,$^{4}$ Aswin P. Vijayan,$^{3}$ Paurush Punyasheel,$^{5}$ \newauthor
William J. Roper,$^{3}$ Louise T. C. Seeyave,$^{3}$ Sonja Soininen,$^{6}$ Peter A. Thomas,$^{3}$\thanks{Deceased} \newauthor Stephen M. Wilkins$^{3}$
\\
$^1$ Key Laboratory for Computational Astrophysics, National Astronomical Observatories, Chinese Academy of Sciences, Beijing 100101, China\\
$^2$ The Institute of Cancer Research, 123 Old Brompton Road, London SW7 3RP, UK\\
$^3$ Astronomy Centre, University of Sussex, Falmer, Brighton BN1 9QH, UK\\
$^4$ Institute of Cosmology and Gravitation, University of Portsmouth, Burnaby Road, Portsmouth PO1 3FX, UK\\
$^5$ Centre for Astrophysics Research, University of Hertfordshire, Hatfield, AL10 9AB, UK\\
$^6$ Department of Physics, University of Helsinki, Gustaf Hällströmin katu 2, FI-00014 Helsinki, Finland\\
}

\begin{document}

\date{Accepted 2025 xxx xx. Received 2025 xxx xx; in original form 2025 xxx xx}

\pagerange{\pageref{firstpage}--\pageref{lastpage}} \pubyear{2025}

\maketitle

\label{firstpage}

\begin{abstract}
The upcoming space-based gravitational wave (GW) observatory, LISA, is expected to detect GW signals from supermassive black hole (SMBH) mergers occurring at high redshifts. However, understanding the origin and growth of SMBHs in the early Universe remains an open problem in astrophysics. In this work, we utilize the First Light And Reionization Epoch Simulations (FLARES), a suite of cosmological hydrodynamical zoom-in simulations, to study SMBH mergers at $5 \la z \la 10$ across a wide range of environments. Most mergers in FLARES involve secondary SMBHs near the seed mass ($m_{\rm seed} \approx 1.5 \times 10^{5}~{\rm M}_{\sun}$) while primary SMBHs span up to $10^{9}~{\rm M}_{\sun}$, resulting in mass ratios from $q \sim 10^{-4}$ to $1$, with a peak at $q \sim 1$. The number of mergers increases rapidly towards lower redshifts, and the comoving total number density scales with overdensity as $n_{\rm merger} = 10^{-3.81} (1 + \delta)^{4.78}$. Denser regions host more massive mergers, with higher merger redshifts and lower mass ratios. Within the FLARES redshift range, LISA is expected to detect mergers with $10^{5} \la M_{\rm tot}/{\rm M}_{\sun} \la 10^{8}$ and $q \ga 10^{-2}$, corresponding to a detection rate of $0.030~{\rm yr}^{-1}$ for events with signal-to-noise ratio ${\rm SNR} \geq 10$. Our study demonstrates the sensitivity of GW predictions at high redshifts to SMBH seed models and merger time delays, highlighting the need for improved modeling in future cosmological simulations to maximize LISA’s scientific return.
\end{abstract}


\section{Introduction}\label{sec:intro}

Although it is well established in modern galaxy formation models that supermassive black holes (SMBHs) play a vital role in the evolution of galaxies \citep[see e.g.][for recent reviews]{Somerville2015,Naab2017,Crain2023}, the origin of SMBHs remains one of the major mysteries in astrophysics \citep{Volonteri2021}. Observations of high-redshift luminous quasars \citep[see][for reviews]{Inayoshi2020,Fan2023} and recent {\it James Webb Space Telescope} (JWST) detections of broad-line active galactic nuclei (AGNs) with moderate to low luminosities at redshifts $5 \la z \la 11$ \citep[also named `little red dots', e.g.][]{Harikane2023,Kocevski2023,Larson2023,Ubler2023,Greene2024,Maiolino2024,Matthee2024} have revealed the presence of SMBHs with masses $10^{6} \la M_{\rm BH}/{\rm M}_{\sun} \la 10^{10}$ in the early Universe. These on-going observations are shaping our understanding of SMBH seed, formation and the co-evolution of SMBHs and galaxies.

JWST data also suggest the possible existence of dual AGNs at $z \ga 4$. For example, \citet{Maiolino2024} identified three dual AGN candidates at $4 \la z \la 6$ from the JADES survey, with projected separations less than ${\sim} 1$ kpc. \citet{Ubler2024} reported an offset AGN at $z = 7.15$ based on JWST/NIRSpec-IFU observations, showing a projected separation of $620$ pc. From the JWST COSMOS-Web survey, \citet{Tanaka2024} found three dual `little red dot' candidates with projected separations of 1--2 kpc at photometric redshifts $z = 6 \sim 7$. \citet{Merida2025} also reported a `little red dot' pair at $z \sim 7$ in the Abell 370 cluster field, with a projected distance of 3.27 kpc, using JWST/NIRCam data from the CANUCS survey. In addition to JWST observations, some quasar pairs with projected separations of ${\sim} 10$ kpc at $z \ga 5$ have recently been reported \citep[see e.g.][]{Yue2021,Yue2023,Matsuoka2024}. These systems may represent merging SMBHs in the early Universe.

During the final stage of their coalescence, two merging SMBHs emit gravitational waves (GWs), which are primary targets of low-frequency GW observatories, including nano-Hertz pulsar timing arrays \citep{Burke-Spolaor2019,Agazie2023,Agazie2023_SMBH,EPTA2023,Reardon2023,Xu2023} and milli-Hertz space-based observatories \citep{Amaro-Seoane2017,Amaro-Seoane2023,Luo2016,Li2024TianQin,Ruan2020}. For example, the space-based Laser Interferometer Space Antenna \citep[LISA,][]{Amaro-Seoane2017,Colpi2024} observatory, planned for launch in the 2030s, will be capable of detecting SMBH mergers with masses $M_{\rm BH} \sim 10^{3}$--$10^{7}~{\rm M}_{\sun}$ up to redshift $z \sim 20$ \citep{Amaro-Seoane2023}. Future LISA observations will offer valuable insights into SMBHs in the early Universe, complementing electromagnetic observations from other telescopes.

To maximize the scientific return of LISA, it is essential to improve our theoretical understanding of its detection targets. Given that SMBHs co-evolve with galaxies \citep[e.g.][]{Kormendy2013}, and LISA will be sensitive to SMBHs across a broad redshift range (i.e. from ${\sim} 0.2$ Gyr after the Big Bang to the present day), studying SMBH mergers in a cosmological framework is critical for LISA forecasts. Cosmological galaxy formation simulations are among the most powerful tools for making theoretical predictions relevant to LISA. 

Several cosmological hydrodynamical simulations with periodic volumes and uniform resolution have been employed to quantify the multi-messenger properties of SMBH mergers and their host galaxies, as well as to predict the detection rates for LISA. These include EAGLE \citep[Evolution and Assembly of GaLaxies and their Environments,][]{Salcido2016}, Illustris \citep{Kelley2017,Katz2020,DeGraf2021,Banks2022}, IllustrisTNG \citep{Li2022,Li2023,DeGraf2021,DeGraf2024}, Horizon-AGN/NewHorizon \citep{Volonteri2020}, ASTRID \citep{Chen2022, DeGraf2024,Wang2025}, and FABLE \citep{Buttigieg2025}. In addition, some zoom-in simulations, which generally have higher numerical resolution than large-volume simulations, have been used to investigate the properties of SMBH mergers and the detectability of their GW signals, see e.g. \citet{Mannerkoski2022,Chakraborty2023,Chakraborty2025,Dong-Paez2023,McCaffrey2025}.

Among the aforementioned uniform-resolution simulations, ASTRID \citep{Ni2022} and TNG300 \citep{Marinacci2018,Naiman2018,Nelson2018,Pillepich2018method,Pillepich2018,Springel2018,Weinberger2017} have the largest volumes, ${\sim}(370~{\rm cMpc})^3$ and ${\sim}(303~{\rm cMpc})^3$, respectively. Other simulations typically have volumes of ${\sim}(100~{\rm cMpc})^3$, which are relatively small for providing robust statistics on high-redshift ($z \ga 5$) SMBH mergers. This is because large samples of high-redshift objects require both a sufficiently large simulation volume to capture rare high-density regions, and high enough resolution to resolve the target objects. On the other hand, despite their higher resolution, zoom-in simulations are often biased toward overdense environments, and are not representative of the cosmic average.

Motivated by the limitations of existing simulations in providing statistically significant samples of SMBH mergers at $z \ga 5$, we adopt the First Light And Reionization Epoch Simulations \citep[FLARES,][]{Lovell2021,Vijayan2021} in this work. FLARES is a suite of zoom-in simulations that resimulate $40$ regions with a wide range of overdensities, selected from a large-volume ($(3.2~{\rm cGpc})^3$) dark matter-only parent simulation, using the EAGLE galaxy formation model \citep{Crain2015,Schaye2015}. By assigning appropriate weights to different regions, FLARES can reproduce the composite statistical properties that match well the cosmic average. Through this novel approach, FLARES achieves a much larger effective simulation volume, and provides a statistically robust sample of extremely massive objects in the densest environments in the early Universe. In addition, FLARES allows us to investigate the environmental dependence of SMBH mergers and their GW signatures.

A detailed investigation of SMBHs and their galaxy hosts in the early Universe ($5 \la z \la 10$) using FLARES is presented in \citet{Wilkins2025}. In addition, \citet{Lovell2023} showed the importance of active SMBHs for producing quenched populations of galaxies at $z \geqslant 5$, providing a good match with early number density constraints in this redshift regime \citep{Carnall2023,Gould2023,Valentino2023}. In this study, we focus on SMBH mergers with FLARES, quantifying their properties, environmental dependence, and implications for LISA detections. In particular, investigating how SMBH merger properties and detection rates depend on environment is a key motivation for this work, enabled by the wide range of environmental overdensities resimulated in FLARES.

The structure of this paper is as follows. In Section~\ref{sec:sim}, we describe the details of FLARES. The properties of FLARES SMBH mergers are quantified in Section~\ref{sec:smbh_prop}, and their environmental dependence is studied in Section~\ref{sec:env_dep}. Section~\ref{sec:gw} is devoted to investigating the GW predictions from FLARES. We summarize and conclude in Section~\ref{sec:summary}.

\section{Simulation details}\label{sec:sim}

\subsection{Overview and weighting scheme}

FLARES \citep{Lovell2021,Vijayan2021} is a suite of 40 zoom-in hydrodynamical simulations performed using the EAGLE galaxy formation subgrid model \citep[see][for details]{Crain2015,Schaye2015}. The spherical zoom-in regions (with a comoving radius of $R_{\rm region} = 14 ~ h^{-1}{\rm cMpc}$), which cover a wide range of overdensities, are selected using the MACSIS simulation volume as a parent \citep[with a comoving box size of $3.2~{\rm cGpc}$,][]{Barnes2017MACSIS}, and they are indexed from 00 to 39. In particular, a number of high-overdensity regions are selected in order to obtain a large sample of the first massive galaxies formed in the Universe. FLARES adopts the cosmological parameters from \citet{Planck2014}, i.e. $\Omega_{\rm m} = 0.307$, $\Omega_{\Lambda} = 0.693$, $\Omega_{\rm b} = 0.04825$, $h = 0.6777$, $\sigma_{8} = 0.8288$, and $n_{\rm s} = 0.9611$. The simulations begin at $z = 127$ and evolve to $z = 4.69$, focusing on the evolution of Universe at high redshifts. The high-resolution dark matter particle mass is $m_{\rm DM} = 9.71 \times 10^{6}~{\rm M}_{\sun}$, and the initial gas particle mass is $m_{\rm gas} = 1.81 \times 10^{6}~{\rm M}_{\sun}$. The gravitational softening length is $2.66~{\rm ckpc}$. Structures, including dark matter haloes and galaxies, are identified using the friends-of-friends \citep[FOF,][]{Davis1985} and SUBFIND \citep{Springel2001} algorithms.

The distribution functions of galaxy properties usually depend on the overdensities of the zoom-in regions. To recover universal distribution functions, FLARES adopts a weighting scheme that assigns a weight to each zoom-in region. These weights are calculated by comparing the mass overdensity distribution in the resimulated volumes to the universal overdensity distribution. Using this weighting scheme, FLARES successfully reproduces composite distribution functions of galaxy properties (e.g. the galaxy stellar mass function) that agree well with observational data and large-volume cosmological simulations \citep{Lovell2021}. For more details, we refer the interested reader to \citet{Lovell2021}, where the weights for all FLARES regions are provided in table A1 (see also the fourth column in Table~\ref{tab:num_mergers}).

\subsection{SMBH subgrid model}\label{subsec:subgrid}

The origins of SMBHs remain an open question. Theoretically, they could form from the remnants of Population III stars, runaway and hierarchical mergers in star clusters, and direct gas collapse \citep[see][for a recent review]{Volonteri2021}. Since the relevant physical scales are unresolved in cosmological simulations, the EAGLE subgrid model seeds an SMBH at the centre of an FOF halo with total mass $M_{\rm FOF} > 10^{10}h^{-1}{\rm M}_{\sun}$ that does not yet contain an SMBH. Specifically, the gas particle with the highest density within the FOF halo is converted into an SMBH particle (i.e. the SMBH seed) with a {\it subgrid} mass of $m_{\rm seed} = 10^{5}h^{-1}{\rm M}_{\sun}$. In the simulations, an SMBH particle has two masses: the subgrid mass, which is used to compute the SMBH accretion rate and increases relatively smoothly, and the {\it dynamical} mass, which governs gravitational interactions and increases only when the SMBH `swallows' a gas particle \citep{Springel2005}. Unless stated otherwise, the SMBH mass hereafter refers to the subgrid mass.

Cosmological simulations often do not accurately model the dynamical friction exerted by other components (e.g. dark matter, gas, and stars) on SMBHs with low masses. To address this, at each time step the EAGLE model repositions SMBHs with masses $M_{\rm BH} < 100\; m_{\rm gas}$ to the minimum of the gravitational potential within the halo. For further details, see \citet{Schaye2015} and for a recent discussion, see \citet{Bahe2022}.

The accretion rate of an SMBH is computed as 
\begin{equation}
    \dot{M}_{\rm accr} = \dot{M}_{\rm Bondi} \times {\rm min}\left(C_{\rm visc}^{-1} (c_{\rm s} / V_{\phi})^3, 1 \right),
\end{equation}
where
\begin{equation}
    \dot{M}_{\rm Bondi} = \frac{4 \upi G^2 M_{\rm BH}^2 \rho}{(c_{\rm s}^2 + v_{\rm rel}^2)^{3/2}}
\end{equation}
is the Bondi accretion rate \citep{Bondi1944}, $G$ is gravitational constant, $\rho$ is gas density, $c_{\rm s}$ is the speed of sound, and $v_{\rm rel}$ is the relative velocity between the SMBH and gas. In addition, the factor $C_{\rm visc}^{-1} (c_{\rm s} / V_{\phi})^3$ is the ratio of the Bondi and the viscous time-scales, where $C_{\rm visc}$ relates to the viscosity of the subgrid accretion disc, and $V_{\phi}$ represents the gas circulation speed at the Bondi radius \citep{Rosas-Guevara2015}. The SMBH accretion rate is further capped by the Eddington rate
\begin{equation}
    \dot{M}_{\rm Edd} = \frac{4 \upi G M_{\rm BH} m_{\rm p}}{\epsilon_{\rm r} \sigma_{\rm T} c},
\end{equation}
where $m_{\rm p}$ is the proton mass, $\epsilon_{\rm r} = 0.1$ is the radiative efficiency, $\sigma_{\rm T}$ is the Thomson cross-section, and $c$ is the speed of light.

The feedback from SMBHs is implemented using a thermal approach, where the energy injection rate from AGN is calculated as 
\begin{equation}
    \dot{E}_{\rm AGN} = \epsilon_{\rm f} \epsilon_{\rm r} \dot{M}_{\rm accr} c^2.
\end{equation}
Here, $\epsilon_{\rm f} = 0.15$ is the feedback efficiency. After each time step $\Delta t$, an amount of feedback energy, $\dot{E}_{\rm AGN} \Delta t$, is added to the energy reservoir of an SMBH. Once the reservoir accumulates sufficient energy, it stochastically releases energy to a neighbouring gas particle as thermal energy, raising the gas temperature by $\Delta T_{\rm AGN}$.

Two SMBHs are merged instantaneously when they satisfy two criteria: (i) their distance is smaller than both the SMBH smoothing length,\footnote{The SMBH smoothing length is defined in the same way as for gas particles in smoothed particle hydrodynamics, using 58 gas neighbours.} $h_{\rm BH}$, and three gravitational softening lengths, and (ii) their relative velocity is smaller than the circular velocity at the radius of $h_{\rm BH}$, i.e. $v_{\rm BH,rel} < \sqrt{G M_{\rm BH} / h_{\rm BH}}$. Note that here, $M_{\rm BH}$ and $h_{\rm BH}$ are the SMBH mass and smoothing length associated with the more massive of the two SMBHs.

Note that FLARES adopts the AGNdT9 parameter configuration from the EAGLE model, which uses a higher value for $C_{\rm visc}$ and $\Delta T_{\rm AGN}$ compared to the reference parameter configuration. Specifically, in the AGNdT9 parameter configuration, $C_{\rm visc} = 2 \upi \times 10^{2}$, $\Delta T_{\rm AGN} = 10^{9}~{\rm K}$. This parameter set better reproduces the properties of hot gas in galaxy groups and clusters, compared with the reference parameters \citep{Barnes2017}. For further details about the EAGLE subgrid model, we refer readers to \citet{Schaye2015,Crain2015}.

\subsection{Retrieving SMBH mergers}

The simulations record the detailed SMBH properties on the fly at every time step when an SMBH is numerically `active' (i.e. when it is updated according to its individual time step bin), saving them in log files. Specifically, when two SMBHs merge, the file saves the merger time (i.e. scale factor), as well as the SMBH IDs and masses. From these files, we can retrieve the information on all SMBH mergers that occurred in FLARES. 

Four low-density FLARES regions (i.e. 27, 29, 30, and 38), which have overdensities $\delta \la -0.1$,\footnote{The overdensity is defined as $\delta \equiv \rho / \bar{\rho} - 1$, where $\bar{\rho}$ is the mean matter density in the Universe and $\rho$ is the smoothed matter density of a FLARES region. Specifically, $\rho$ is computed by first constructing a density field using the nearest grid point mass assignment scheme, then smoothing it using a top-hat kernel with a smoothing scale of $14~h^{-1}{\rm cMpc}$ (the radius of each FLARES region). The smoothed density at the centre of each region is denoted as $\rho$. See \citet{Lovell2021} for details.} do not have any SMBH mergers. In total, 2,017 SMBH mergers are identified across the remaining 36 FLARES regions (see Table~\ref{tab:num_mergers} for details). For comparison, within the same redshift range, the EAGLE Ref-L100N1054 run, with a comoving box size of $100~{\rm cMpc}$ and identical resolution, contains only 182 SMBH mergers. This pronounced difference highlights the capability of FLARES -- which includes high-density regions that are not well sampled in cosmological simulations with relatively small volumes, but precisely where high-redshift structures predominantly form -- to study SMBH merger events in the early Universe ($z \ga 5$).

\begin{table}
\centering
\caption{Summary of FLARES regions. From left to right, the columns show the region index, the number of SMBH mergers ($N_{\rm merger}$), the overdensity ($\delta$), and the weight ($w$). The regions are sorted by their overdensity in descending order.}
\label{tab:num_mergers}
\begin{tabular}{cccc}
\hline
Region index & $N_{\rm merger}$ & $\delta$ & $w$ \\
\hline
00 & 114 & 0.970 & 0.000 027 \\
01 & 83 &  0.918 & 0.000 196 \\
02 & 94 &  0.852 & 0.000 429 \\
03 & 105 & 0.849 & 0.000 953 \\
04 & 117 & 0.846 & 0.000 444 \\
05 & 101 & 0.842 & 0.000 828  \\
06 & 124 & 0.841 & 0.000 666 \\
07 & 109 & 0.839 & 0.001 178 \\
08 & 179 & 0.839 & 0.000 265 \\
09 & 66 &  0.833 & 0.001 029 \\
10 & 108 & 0.830 & 0.000 387 \\
11 & 97 &  0.829 & 0.000 719 \\
12 & 98 &  0.828 & 0.000 668 \\
13 & 102 & 0.824 & 0.000 488 \\
14 & 133 & 0.821 & 0.001 190 \\
15 & 110 & 0.820 & 0.000 757 \\
16 & 55 &  0.616 & 0.003 738 \\
17 & 47 &  0.616 & 0.004 678 \\
18 & 37 &  0.431 & 0.009 359 \\
19 & 36 &  0.431 & 0.012 324 \\
21 & 7 &   0.266 & 0.027 954 \\
20 & 17 &  0.266 & 0.029 311 \\
22 & 25 &  0.121 & 0.057 876 \\
23 & 5 &   0.121 & 0.062 009 \\
36 & 13 &  0.055 & 0.070 408 \\
37 & 5 &   0.055 & 0.062 451 \\
24 & 6 &  $-$0.007 & 0.074 502\\
25 & 2 &  $-$0.007 & 0.080 377 \\
35 & 4 &  $-$0.007 & 0.076 486 \\
34 & 4 &  $-$0.007 & 0.076 001 \\
32 & 3 &  $-$0.066 & 0.064 280 \\
33 & 2 &  $-$0.066 & 0.066 277 \\
26 & 2 &  $-$0.121 & 0.063 528 \\
27 & 0 &  $-$0.121 & 0.058 231 \\
28 & 4 &  $-$0.222 & 0.034 467 \\
29 & 0 &  $-$0.222 & 0.024 216 \\
31 & 2 &  $-$0.311 & 0.013 127 \\
30 & 0 &  $-$0.311 & 0.012 087 \\
39 & 1 &  $-$0.434 & 0.003 366 \\
38 & 0 &  $-$0.479 & 0.002 721 \\
\hline
\end{tabular}
\end{table}

\section{Properties of high-redshift SMBH mergers} \label{sec:smbh_prop}

The masses (and mass ratios), spins, and merger redshifts of SMBHs are crucial inputs in calculating the GW signals from SMBH mergers. These quantities from cosmological simulations are particularly useful in predicting GW event rates and the stochastic GW background for future space-based GW observatories. Although FLARES does not explicitly model the evolution of SMBH spins,\footnote{To model the evolution of SMBH spins in galaxy formation simulations, it is necessary to construct an accretion disc subgrid model that captures the transfer of angular momentum from the inflowing gas to the SMBH. See, for example, \citet{Dubois2014,Fiacconi2018,Bustamante2019,Cenci2021,Talbot2021,Husko2022,Koudmani2024,Sala2024} for some recent progresses in this research direction.} it offers valuable insights into the statistics of SMBH masses, mass ratios, and merger redshifts in the early Universe. In this section, we quantify the properties of SMBH mergers from FLARES.

\subsection{Component masses}

\begin{figure} 
\centering\includegraphics[width=\columnwidth]{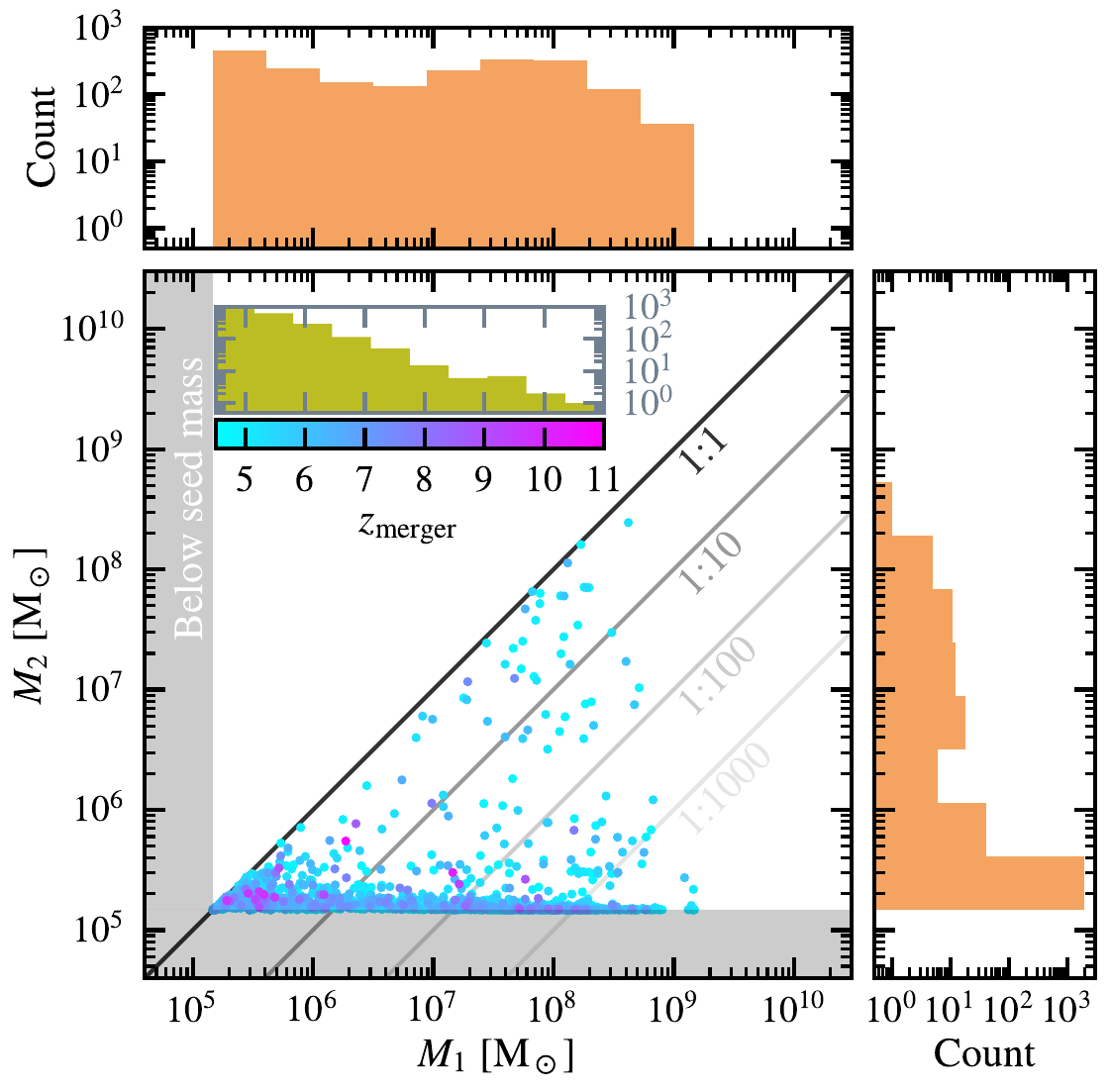}
\caption{Primary ($M_{1}$) and secondary ($M_{2}$) SMBH masses of all SMBH mergers in FLARES. Each dot represents an SMBH merger and it is colour-coded by the redshift of the event, $z_{\rm merger}$. The solid lines show the mass ratios ($M_{2}$:$M_{1}$) from 1:1 to 1:1000. The mass range below the SMBH seed mass ($10^5 h^{-1}{\rm M}_{\sun}$) is marked by a grey colour. The top and right sub-panels show the histograms of $M_{1}$ and $M_{2}$, respectively. The inset panel displays the histogram of $z_{\rm merger}$. The majority of the SMBH mergers occur at $z \la 7$ and have their $M_{2}$ close to the seed mass.}
\label{fig:m1_vs_m2}
\end{figure}

\begin{figure*} 
\centering\includegraphics[width=2.05\columnwidth]{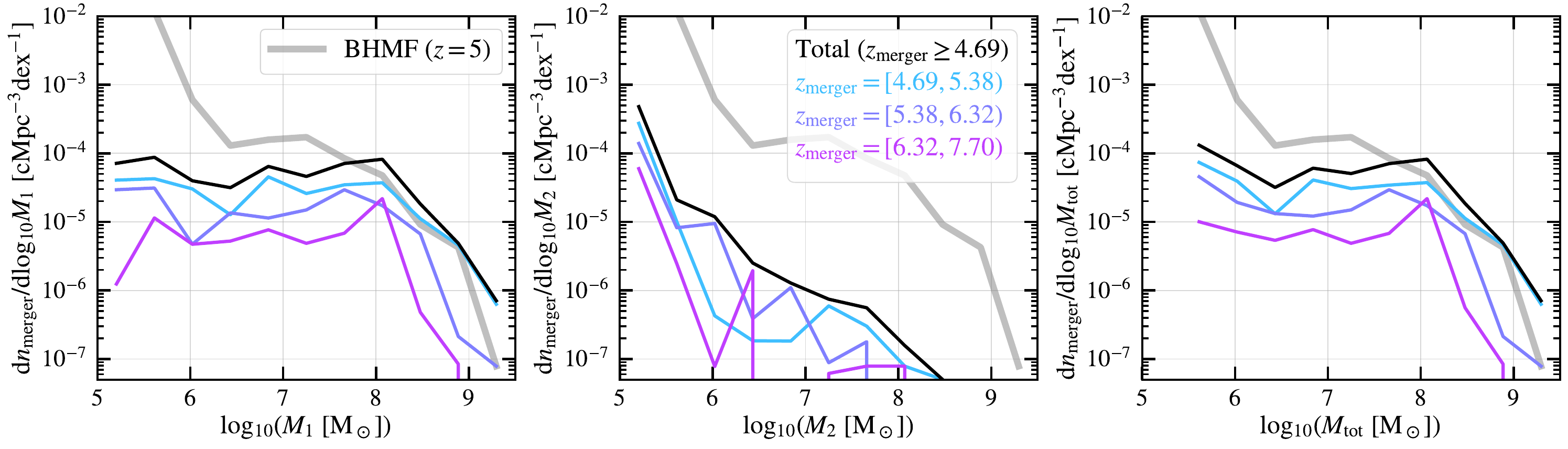}
\caption{Composite mass functions of SMBH mergers for the primary SMBH ($M_1$, left), the secondary SMBH ($M_2$, middle), and the SMBH binary ($M_{\rm tot} = M_1 + M_2$, right). Different merger redshift ranges are plotted using different colours, as specified in the legend. The redshift bins are chosen such that each spans an equal time interval of $200$ Myr. The total mass functions, which account for SMBH mergers across all redshifts in FLARES, are shown as black curves. For comparison, the composite black hole mass function -- including all SMBHs, not just those that have experienced mergers -- at $z = 5$ is plotted as a gray line in each panel.  The secondary SMBHs are dominated by those with masses close to the seed mass,  particularly at high redshifts.}
\label{fig:bh_mass_func}
\end{figure*}

\begin{figure*} 
\centering\includegraphics[width=2.05\columnwidth]{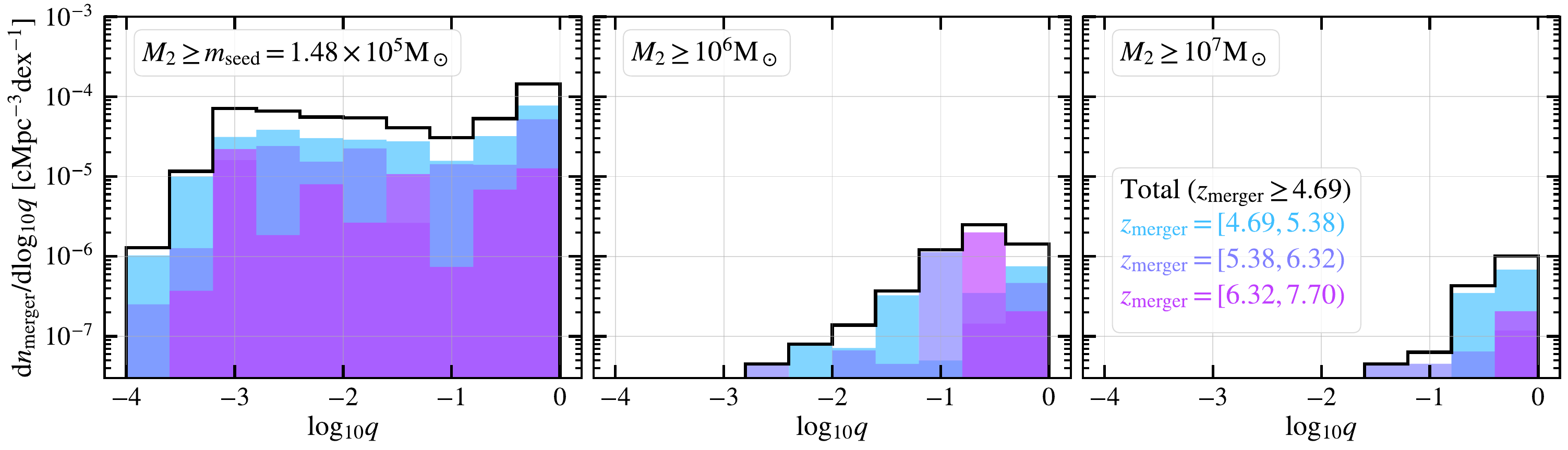}
\caption{Composite distributions of the SMBH merger mass ratio $q = M_2 / M_1$. The panels, from left to right, show distributions with different cuts in $M_2$: $M_{2} \geq m_{\rm seed} = 1.48 \times 10^{5}~{\rm M}_{\sun}$, $M_{2} \geq 10^{6}~{\rm M}_{\sun}$, and $M_{2} \geq 10^{7}~{\rm M}_{\sun}$. In each panel, the total distribution is shown in black, while the distributions of SMBH mergers at different $z_{\rm merger}$ ranges are plotted in different colours, as specified in the legend. For all SMBH mergers in FLARES (black composite histogram in the left-hand panel), the mass ratios span a wide range, from ${\sim}10^{-4}$ to $1$.}
\label{fig:mass_ratio}
\end{figure*}

In Figure~\ref{fig:m1_vs_m2}, we show the primary ($M_{1}$) and secondary ($M_{2}$) SMBH masses of the 2,017 SMBH mergers retrieved from all FLARES regions, with each dot colour-coded by its merger redshift, $z_{\rm merger}$. The primary SMBH is defined as the more massive one in a merger (i.e. $M_{1} \geq M_{2}$), therefore, all data points locate below the 1:1 line. Most SMBH mergers (${\sim} 96 \%$) occur at $z \la 7$. While the histogram of primary SMBH masses is relatively flat across the range $[m_{\rm seed}, {\sim}10^{8}~{\rm M}_{\sun}]$, the secondary SMBH masses are strongly clustered around $m_{\rm seed}$. This indicates that the majority of SMBH mergers in FLARES involve an SMBH with a mass close to the seed mass. 

Figure~\ref{fig:m1_vs_m2} presents only the histograms of SMBH masses. To derive the {\it composite} distribution of SMBH masses, which represents the average distribution in the Universe, we apply the FLARES weighting scheme. Specifically, the mass function is first computed for each region, then multiplied by its corresponding region weight, and finally summed across all regions to obtain the composition function.

The composite SMBH mass functions of primary and secondary SMBHs are plotted as black curves in the left and middle panels of Figure~\ref{fig:bh_mass_func}, respectively. Additionally, the mass functions of SMBH mergers within different redshift ranges are also presented. To enable direct comparison among different redshift bins, we have chosen the bins so that each spans an equal time interval of $200$ Myr. The composite mass functions support our observations from Figure~\ref{fig:m1_vs_m2}: primary SMBHs tend to have a flat distribution at $M_{1} \la 10^{8}~{\rm M}_{\sun}$, followed by a steep decline at higher masses. This behaviour remains qualitatively similar across different redshift ranges, although the lower redshift bins tend to show a higher number of SMBH mergers. In contrast, the mass function of secondary SMBHs peaks near $m_{\rm seed}$ and decreases sharply at higher masses. This trend is particularly pronounced at higher redshifts. For example, nearly all secondary SMBHs with $z_{\rm merger} \ga 6.32$ have $M_{2} \la 10^{6}~{\rm M}_{\sun}$. 

Figure~\ref{fig:bh_mass_func} also shows the composite black hole mass function (BHMF) for {\it all} SMBHs at $z = 5$ (gray line), allowing for comparison with the primary/secondary mass functions discussed above for SMBHs that have experienced mergers \citep[for a detailed study of the evolution of the BHMF in FLARES, see][]{Wilkins2025}. This BHMF exhibits a similar or lower SMBH abundance at the high-mass end compared to the primary mass function of SMBH mergers (left panel), reflecting the fact that most of these high-mass SMBHs have undergone multiple mergers at higher redshifts. However, the BHMF at $z = 5$ contains significantly more low-mass SMBHs (i.e. $\ga$ two orders of magnitude higher) that have not experienced mergers. Note that similar behaviour is also observed in the TNG300 and ASTRID simulations; see figure 4 of \citet{DeGraf2024} for details.

In the right panel, we show the composite mass function of the total SMBH masses (i.e. $M_{\rm tot} = M_{1} + M_{2}$). Since most secondary SMBHs have masses close to the seed mass, the composite total mass functions closely resemble those of primary SMBHs, except at the low-mass end. 

\subsection{Mass ratios}\label{subsec:mass_ratio}

The distinct mass functions of primary and secondary SMBHs significantly influence the distribution of mass ratios, which, in turn, will affect the predicted GW signals. In Figure~\ref{fig:mass_ratio}, we plot the composite distributions of SMBH mass ratios, 
\begin{equation}
    q \equiv \frac{M_{2}}{M_{1}}.
\end{equation}
To better understand the dependence on merger redshifts, we also show the $q$-distributions for mergers with different $z_{\rm merger}$ ranges, similar to Figure~\ref{fig:bh_mass_func}. Furthermore, we plot the distributions for different secondary SMBH mass ranges (from left to right panels), examining the $q$-distributions for more massive versus less massive SMBH mergers.

For all SMBH mergers in FLARES (represented by the black composite histogram in the left-hand panel), the mass ratios span a wide range, from ${\sim}10^{-4}$ to $1$. The distribution peaks at $q \sim 1$ and remains relatively flat down to $q \sim 10^{-3}$. Note that, apart from massive, closely equal-mass SMBH mergers, the $q \sim 1$ bin also contains mergers in which both SMBHs have masses close to the seed mass. The lowest mass ratios (i.e. $q \la 10^{-3}$) occur when the most massive SMBHs in the simulations merge with seed-mass SMBHs. This explains why mergers at higher $z_{\rm merger}$ (represented by the darker-coloured histograms) have larger minimum $q$ values, as they are constrained by the masses of the most massive SMBH present at those redshifts.

Focusing on SMBH mergers with higher component masses (e.g. $M_{2} \geq 10^{6}~{\rm M}_{\sun}$ in the middle panel and $M_{2} \geq 10^{7}~{\rm M}_{\sun}$ in the right-hand panel), which effectively excludes mergers involving seed-mass SMBHs, we find significantly different $q$-distributions. These distributions peak around $q \sim 1$ and decrease monotonically toward lower $q$. Similarly to previous findings, mergers at higher $z_{\rm merger}$ exhibit larger $q$ values, as the distributions are constrained by the masses of the most massive SMBHs existing at those redshifts.

Note that the aforementioned distributions of SMBH masses and mass ratios could be affected by the lack of accurate dynamical treatment in FLARES. First, low-mass SMBHs ($M_{\rm BH} < 100\; m_{\rm gas}$) are manually repositioned to the local minimum of the gravitational potential at each time step. This procedure could overestimate dynamical friction and artificially enhance the sinking and merger efficiency of low-mass SMBHs \citep[e.g.][]{Tremmel2015}, contributing to the peak near $m_{\rm seed}$ in the $M_{2}$ distribution and to the mass ratio distribution in the low-$q$ regime. Several recent galactic-scale simulations which better resolve the dynamical evolution of low-mass SMBHs ($\la 10^{5}~{\rm M}_{\sun}$) have demonstrated the challenges of SMBH seed sinking, especially in high-redshift, clumpy galaxies \citep[e.g.][]{Tremmel2018,Pfister2019,Ma2021,Partmann2024,Zhou2025}. Second, in FLARES, two SMBHs are merged instantaneously when their separation becomes smaller than the spatial resolution (typically at kpc or sub-kpc scales) and their relative velocity is low. As a result, the dynamical evolution below $\sim$kpc scales is unresolved. Some previous studies have included post-processing delay time models to account for the time-scale below the numerical resolution and have found that the merger efficiency of low-$M_{\rm tot}$ and low-$q$ systems can be significantly affected \citep[e.g.][]{Katz2020,Volonteri2020,Chen2022}. In addition, during the delay time, the infalling SMBHs can continue to grow in mass, leading to higher merger masses and different mass ratios when they actually coalesce \citep[see e.g.][]{Banks2022,Liao2023}. Therefore, the distributions presented in Figures \ref{fig:m1_vs_m2}--\ref{fig:mass_ratio} could differ substantially if dynamical evolution were treated more accurately, which should be better quantified in future simulations that include improved SMBH dynamics.

\subsection{Merger redshifts}

\begin{figure} 
\centering\includegraphics[width=\columnwidth]{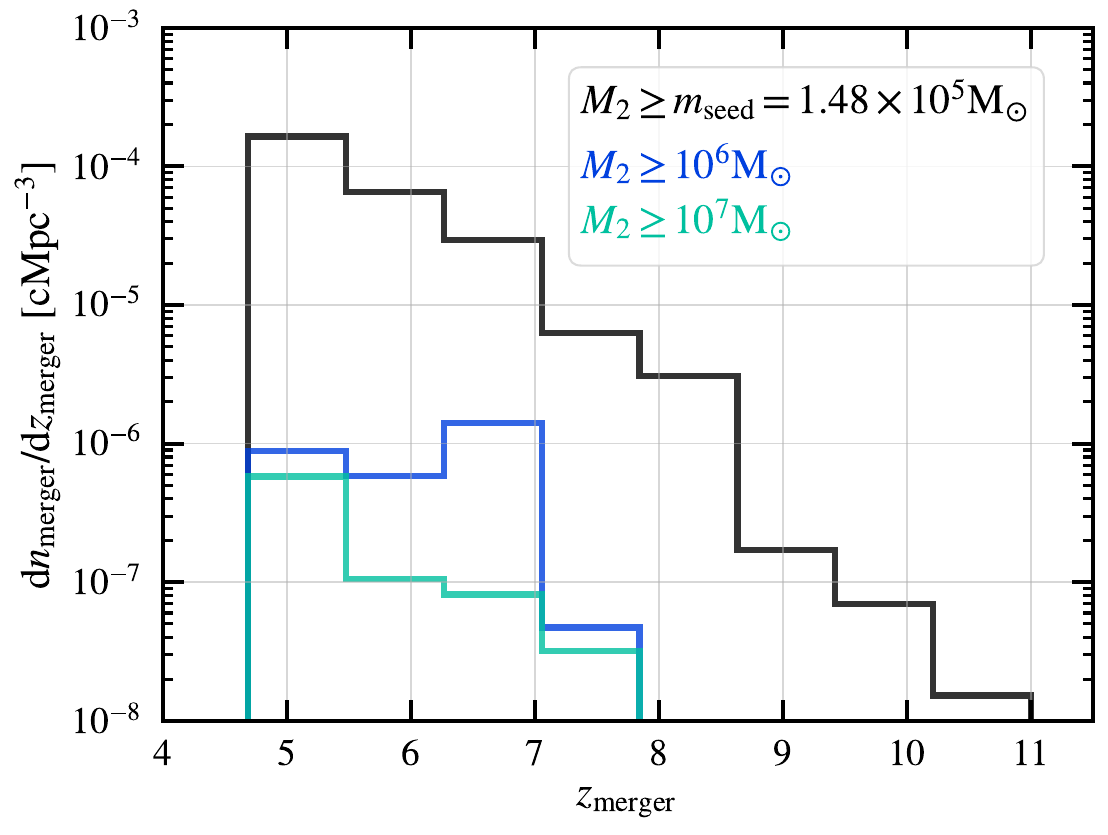}
\caption{Composite distribution of merger redshifts. As indicated in the legend, the distribution of all mergers (i.e. $M_{2} \geq m_{\rm seed}$) is plotted in black, while the distributions of SMBH mergers with different $M_{2}$ cuts are shown in different colours. The merger number density increases rapidly as redshift decreases.}
\label{fig:num_dens_z}
\end{figure}

The merger redshift of an SMBH binary is a key parameter in calculating GW signals, as it determines how the GW frequency is redshifted when observed by our instruments and affects the observed amplitude. The composite distribution of merger redshifts is shown in Figure~\ref{fig:num_dens_z} (the black line). As expected, the distribution function increases monotonically toward lower redshifts, reflecting the higher frequency of SMBH mergers at later times due to the rising number of galaxy mergers. For higher cuts in $M_{2}$ (the lines with other colours), the distributions skew toward lower redshifts, as these mergers involve more massive SMBHs.

In FLARES, two SMBHs are merged instantaneously when they satisfy the distance and velocity criteria mentioned in Section~\ref{subsec:subgrid}. This ad hoc implementation overestimates the actual merger redshift of an SMBH binary, as the true merger redshift should be lower than that provided in the simulations. The difference between the merger redshift in cosmological hydrodynamical simulations and the actual merger redshift remains an open question, involving a series of complicated physical processes \citep[see e.g.][for a review]{Amaro-Seoane2023}. This difference is expected to depend on the properties of the merging galaxies. In Section~\ref{sec:gw}, we explore the impact of this uncertainty on GW predictions.

\subsection{Impact of SMBH seed mass} \label{subsec:bh_seed_prop}

\begin{figure*} 
\centering\includegraphics[width=2\columnwidth]{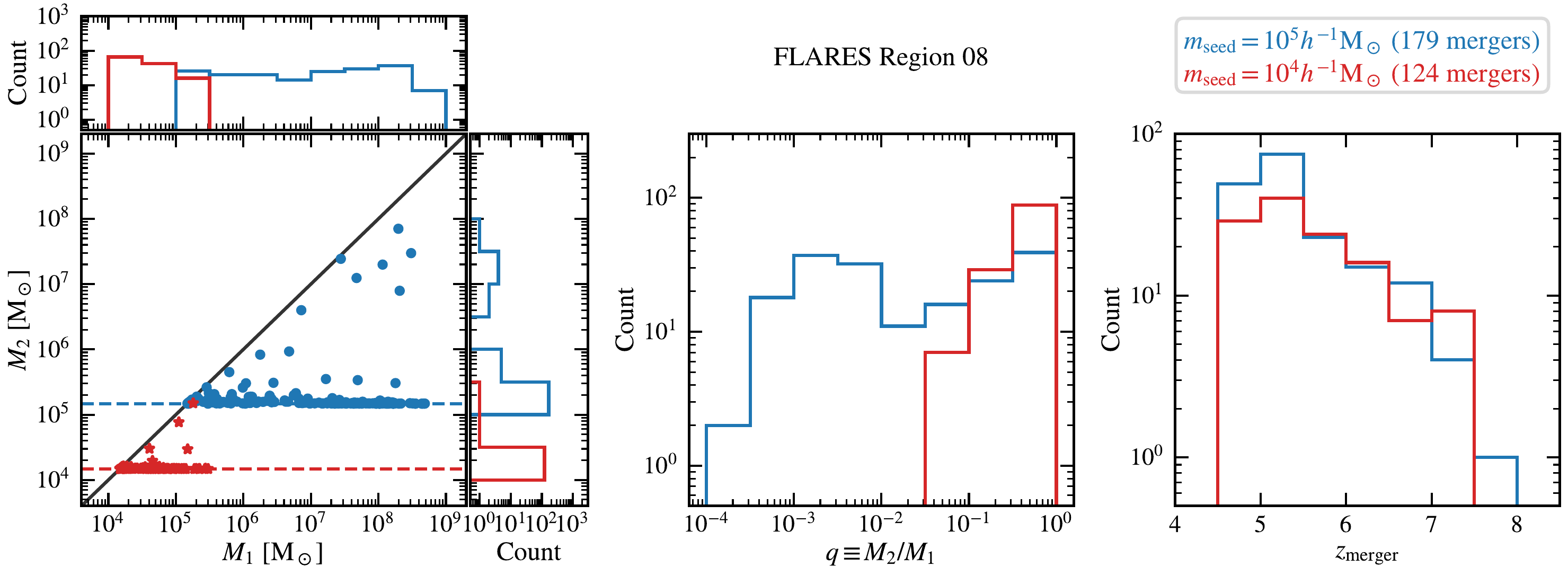}
\caption{Tests on the impact of the SMBH seed mass using Region 08. The fiducial run with $m_{\rm seed} = 10^{5} ~ h^{-1}{\rm M}_{\sun}$ is plotted in blue, while the test run with a lower SMBH seed mass (i.e. $m_{\rm seed} = 10^{4} ~ h^{-1}{\rm M}_{\sun}$) is shown in red. {\it Left-hand panel:} The secondary SMBH mass versus the primary SMBH mass of all mergers in Region 08. The diagonal solid line marks the mass ratio of $1:1$, and the two horizontal dashed lines indicate the seed masses. The right and top sub-panels compare the histograms from two runs. {\it Middle panel:} The histograms of the mass ratios for all merger events. {\it Right-hand panel:} The histograms of the merger redshifts from the two runs. Compared to the fiducial run, the test run with a lower SMBH seed mass exhibits fewer SMBH mergers, fails to produce massive SMBHs (i.e. $M_{\rm BH} \ga 10^{6}~{\rm M}_{\sun}$), has fewer SMBH mergers at low redshifts, and has more mergers with high mass ratios.}
\label{fig:m_seed_test}
\end{figure*}

The above discussions demonstrate that SMBH masses (particularly the secondary SMBH mass) and mass ratios are strongly influenced by the SMBH seed mass adopted in simulations. Since the formation of SMBH seeds is not yet fully understood, the seeding mechanism in current simulations is significantly simplified.

To investigate the impact of the SMBH seed mass, we reran Region 08, which has the highest number of SMBH mergers in the fiducial FLARES suite (179 mergers in total). In this test run, all parameters were kept unchanged except for the seed mass, which was set to a lower value, $m_{\rm seed} = 10^{4} ~ h^{-1}{\rm M}_{\sun}$. In this test run, the total number of SMBH mergers is lower, with only 124 mergers occurring. In Figure~\ref{fig:m_seed_test}, we compare the distributions of SMBH masses, mass ratios, and merger redshifts between the fiducial and test runs. 

As shown in the left-hand panel, the test run produces significantly lower primary and secondary SMBH masses compared to the fiducial run. Both $M_{1}$ and $M_{2}$ peak at $m_{\rm seed}$, and they are all below $10^{6}~{\rm M}_{\sun}$. With lighter SMBH seeds, an SMBH must undergo more merger events to achieve a certain mass. Additionally, in the Bondi accretion model, a lower SMBH mass results in reduced accretion rates, further slowing SMBH growth during the early stages. Furthermore, lighter SMBHs have lower circular velocities at $h_{\rm BH}$ (since $v_{\rm circ} \propto \sqrt{M_{\rm BH}}$), which requires a lower relative velocity for SMBHs to merge. This could make certain merger configurations more difficult to satisfy the merger criteria. These factors collectively explain the lower SMBH masses observed in the test run.

Due to the absence of high-mass SMBHs (i.e. $M_{\rm BH} \ga 10^{6}~{\rm M}_{\sun}$) and the narrower distribution of SMBH masses in the test run, the mass ratios span a more restricted range (i.e. from ${\sim} 0.01$ to $1$) and peak at $q = 1$, as plotted in the middle panel. For the merger redshifts, presented in the right-hand panel, the test run exhibits a similar distribution to the fiducial run at $z_{\rm merger} \ga 5.5$ but it shows a significantly lower number of SMBH mergers at $z \la 5.5$.

Whilst this test provides a critical assessment of the impact of seed mass on merger property distributions, it is worth keeping in mind that many parameters of the EAGLE model were calibrated to distribution functions at $z = 0.1$ whilst explicitly assuming the fixed higher seed mass.
Changing the seed mass can therefore lead to complicated covariances with other parameters considered in the calibration procedure \citep[see][]{Villaescusa-Navarro2021,Kugel2023}, leading to knock on effects on SMBH merger and GW event properties.

The above comparisons between the test run and the fiducial run highlight the critical impact of the SMBH seed mass on SMBH merger properties. On one hand, this result implies that we need a more sophisticated SMBH seeding model in next-generation cosmological hydrodynamical simulations \citep[see e.g.][for recent work]{Taylor2014,Wang2019,DeGraf2020,Bhowmick2022,Bhowmick2024a,Bhowmick2024b,Bhowmick2024c} to provide better theoretical predictions. On the other hand, it also suggests that the GW signals from SMBH mergers in the early Universe can provide stringent constraints on our understanding of the origin of SMBHs at high redshifts.

\section{Dependence on environments} \label{sec:env_dep}

\begin{figure} 
\centering\includegraphics[width=\columnwidth]{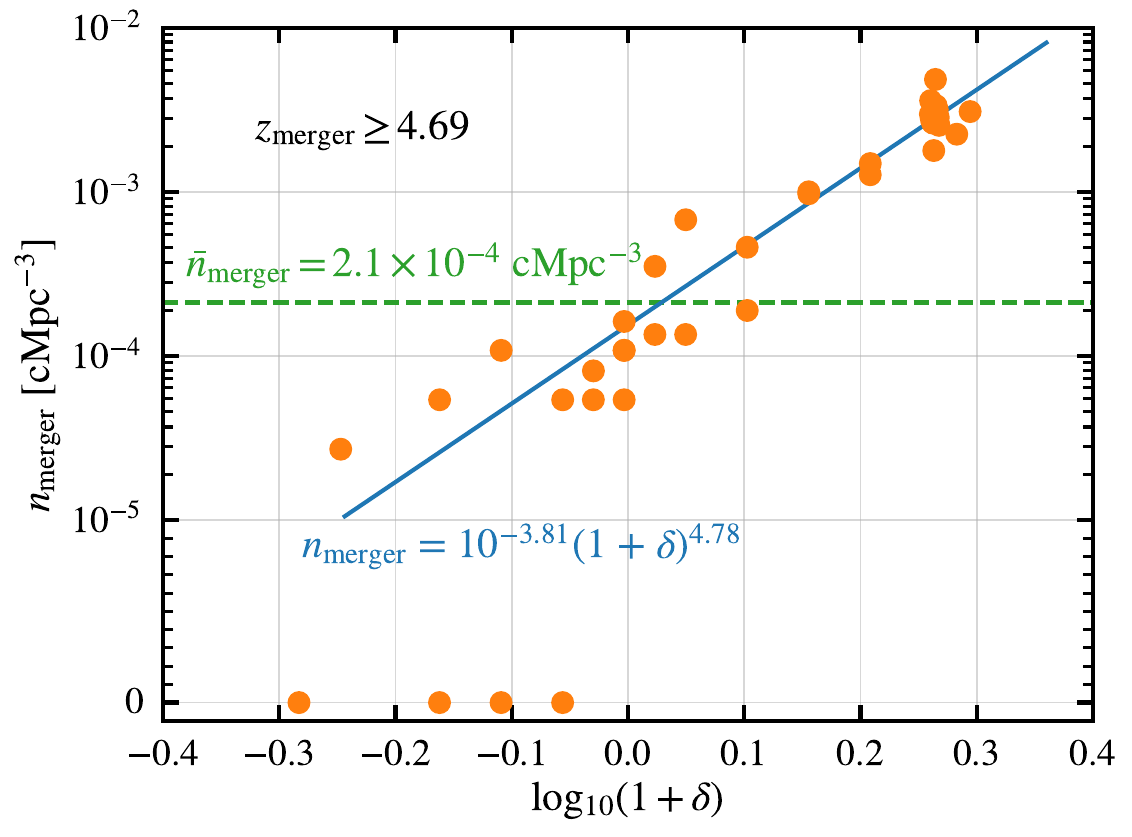}
\caption{Environmental dependence of SMBH merger number density. Note that four FLARES regions (i.e. Regions 27, 29, 30, and 38) have zero mergers. The blue line shows the best-fitting relation between the merger number density and the overdensity of FLARES regions, obtained using Bayesian inference with a Poisson likelihood (see Appendix~\ref{ap:fitting} for details). The composite merger number density computed from all 40 regions, $\bar{n}_{\rm merger} = 2.1 \times 10^{-4}~{\rm cMpc}^{-3}$, is indicated by the dashed horizontal line. Note that for clarity, the $y$-axis uses a linear scale in the range $[0, 10^{-5})$ and a logarithmic scale for $n_{\rm merger} \geq 10^{-5}~{\rm cMpc^{-3}}$. As expected, SMBH merger events strongly depend on environmental overdensities.}
\label{fig:num_dens_env}
\end{figure}

\begin{figure*} 
\centering\includegraphics[width=2\columnwidth]{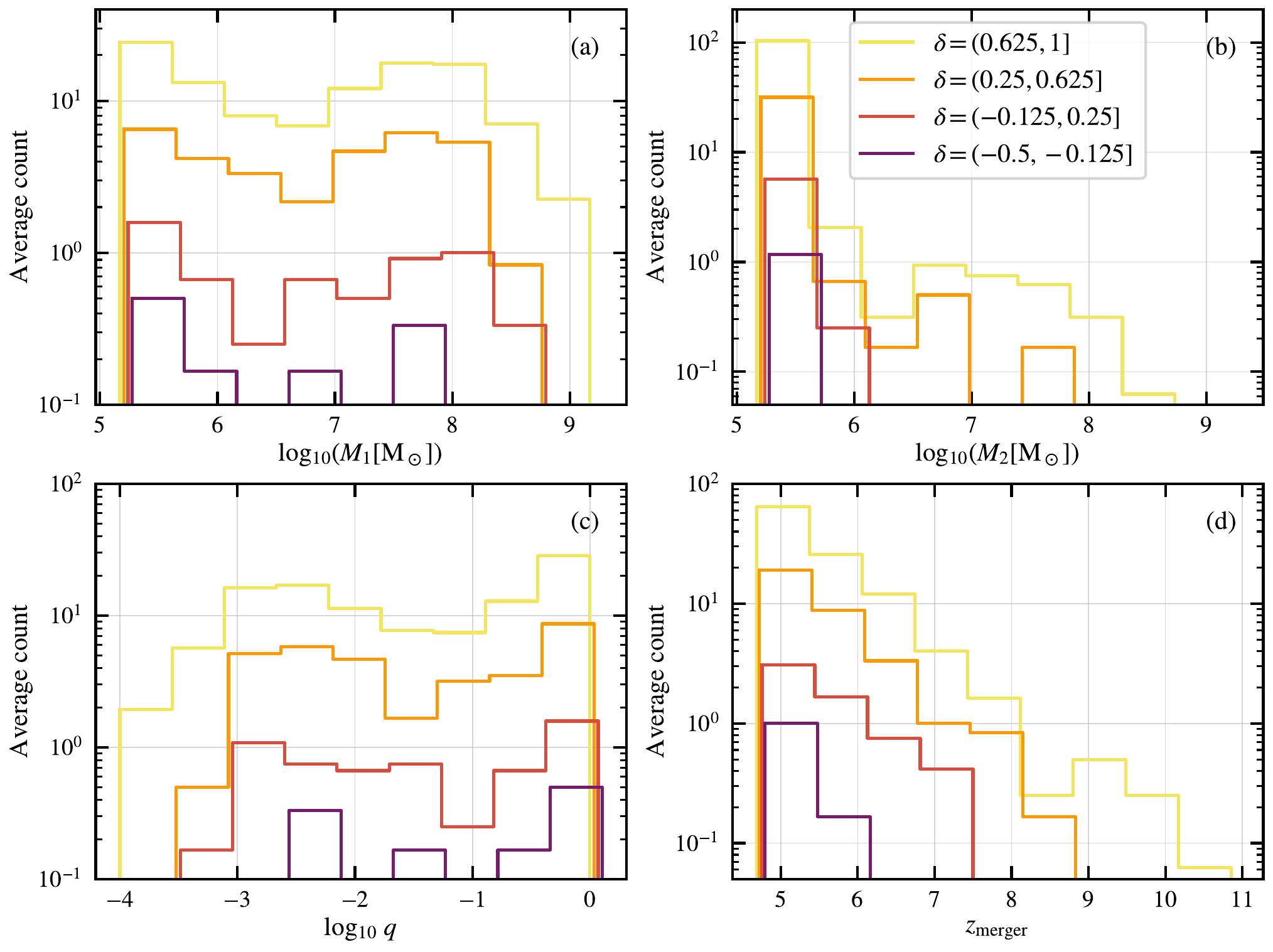}
\caption{Environmental dependence of the properties of SMBH mergers. Panels (a) to (d) show the average histograms of primary SMBH masses ($M_{1}$), secondary SMBH masses ($M_{2}$), mass ratios ($q$), and merger redshifts ($z_{\rm merger}$), respectively. The FLARES regions are grouped into four bins based on their overdensities, and average histograms are computed from all regions within each overdensity bin. The colour conventions for the different $\delta$ bins are given in the legend of Panel (b). To enhance clarity, histograms corresponding to different bins are offset slightly along the horizontal axis. The regions with higher environmental overdensity tend to form more massive SMBHs, exhibit lower mass ratios, and have SMBH mergers occurring earlier, while the regions with lower overdensity show the opposite trends.}
\label{fig:env_impact}
\end{figure*}

FLARES consists of 40 regions that span a wide range of environmental overdensities, providing a unique framework to investigate the dependence of galaxy and SMBH properties on their surroundings. The environmental dependence of galaxy stellar mass functions, star formation rate distribution function, and black hole mass functions has been explored in previous FLARES studies \citep{Lovell2021,Wilkins2025}. In this section, we further examine how SMBH merger properties depend on environments.

In Figure~\ref{fig:num_dens_env}, we plot the comoving number densities of SMBH mergers, $n_{\rm merger} = N_{\rm merger} / V_{\rm c}$, from different regions as a function of overdensity, $\delta$. Here, $V_{\rm c} = 4 \pi R_{\rm region}^3 / 3$ is the comoving volume of a FLARES region, and the overdensities of different regions are listed in the third column of Table~\ref{tab:num_mergers}. As expected, denser regions tend to have a higher number of SMBH mergers. This dependence is well described by a power-law relation, $n_{\rm merger} = 10^{-3.81} (1 + \delta)^{4.78}$ (see Appendix~\ref{ap:fitting} for details). Within the overdensity range covered by FLARES, i.e. $-0.3 \la \log_{10}(1 + \delta) \la 0.3$ (or equivalently $-0.5 \la \delta \la 1$), the number density in the densest region is ${\sim}500$ times that in the least dense region. 

The environmental dependence of other merger properties, including SMBH masses, mass ratios, and merger redshifts, is summarized in Figure~\ref{fig:env_impact}. In this figure, the FLARES regions are grouped into four bins based on their overdensities, and the average histograms for the different $\delta$ bins are plotted. Lighter (darker) colours represent denser (less dense) regions. These histograms reveal that denser environments tend to host more massive SMBHs, particularly for secondary SMBHs. For example, SMBHs with $M_{2} \ga 10^{6}~{\rm M}_{\sun}$ appear only in regions with $\delta \ga 0.25$. Furthermore, SMBH mergers in denser environments exhibit lower mass ratios due to the presence of more massive primary SMBHs. Finally, denser regions tend to host mergers at higher redshifts, reflecting the earlier formation and growth of SMBHs in these environments.

The results in this section suggest that low-frequency GW signals from high-redshift SMBHs are primarily dominated by sources in overdense regions. The environmental dependence of SMBH merger properties revealed in this section provides useful context for previous studies that use cosmological zoom-in simulations to investigate SMBH mergers \citep[e.g.][]{Mannerkoski2022,Dong-Paez2023,McCaffrey2025}. These zoom-in simulations typically select overdense regions for resimulation, which may introduce biases in the quantification of SMBH properties. By applying the quantified environmental dependence of merger number density shown in Figure~\ref{fig:num_dens_env}, and knowing the overdensity of a given zoom-in region, one can roughly estimate the difference in merger number densities between that region and a region more representative of the cosmic mean (i.e. $\delta \approx 0$ or the composite merger number density). In this way, the broader environmental insights provided by FLARES can aid in contextualizing and interpreting results from targeted zoom-in simulations.

\section{Implications on GW observations} \label{sec:gw}

\begin{figure*} 
\centering\includegraphics[width=2\columnwidth]{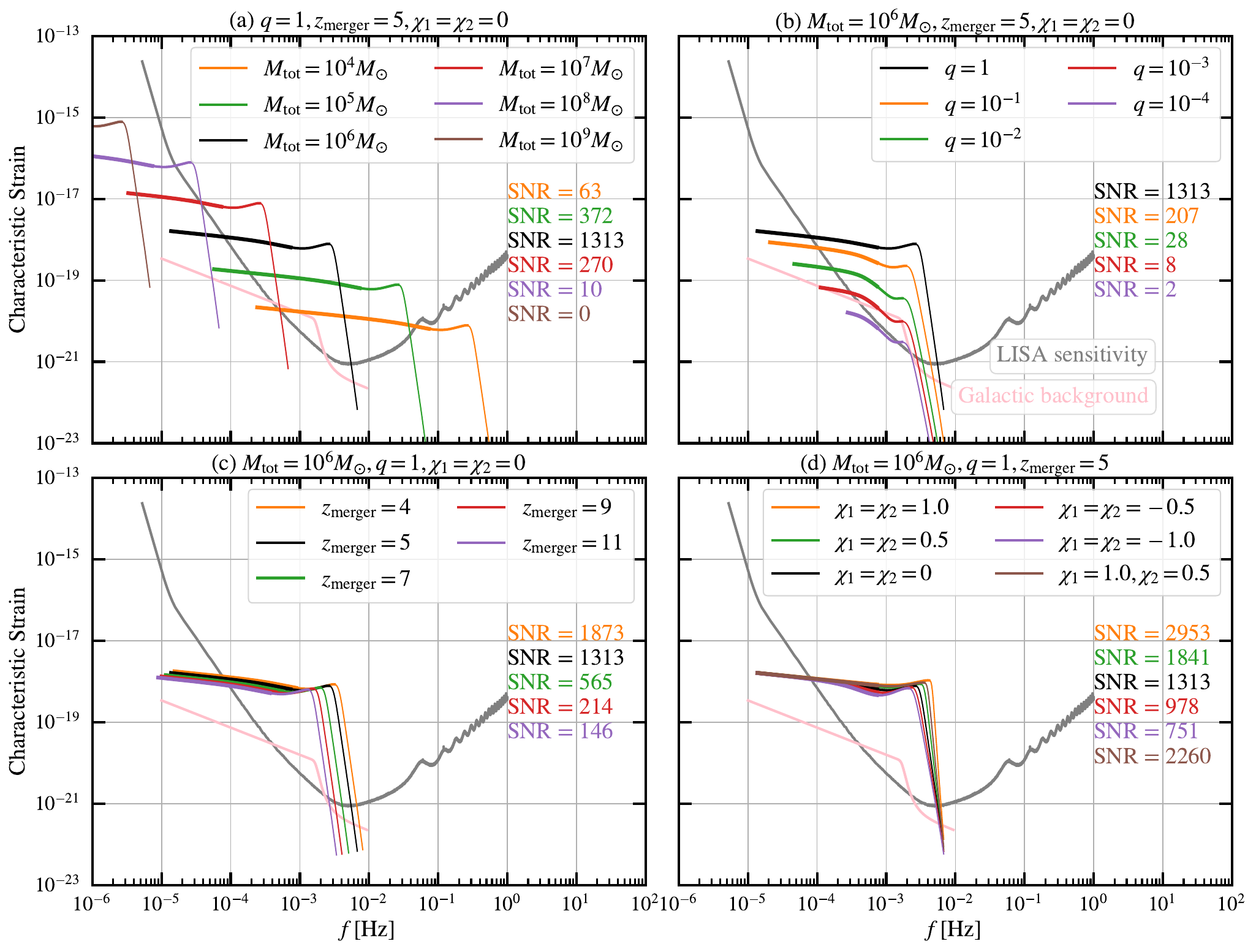}
\caption{Dependence of GW signals on SMBH merger properties. Panels (a) to (d) show the dependence on total SMBH mass ($M_{\rm tot}$), mass ratio ($q$), merger redshift ($z_{\rm merger}$), and SMBH spins ($\chi_1$ and $\chi_2$), respectively, while fixing the values specified at the top of each panel. In each panel, the gray curve represents the LISA sensitivity curve, while the pink curve shows the Galactic white dwarf background \citep{Bender1997,Hiscock2000}. The characteristic strains for different SMBH merger parameters are distinguished by colours, with the case of $M_{\rm tot} = 10^{6}~{\rm M}_{\sun}$, $q = 1$, $z_{\rm merger} = 5$, and $\chi_1 = \chi_2 = 0$ plotted in black and included in all panels for reference. In each GW signal, the thickest, thinner, and thinnest line segments represent the inspiral, merger, and ringdown phases, respectively, progressing from low to high frequencies. The SNR values, calculated including the Galactic background noise, are displayed in each panel. In the range of plotting parameters, the total SMBH mass and mass ratio have a more significant impact on GW signals compared to merger redshifts and SMBH spins.}
\label{fig:gw_dep}
\end{figure*}

Given that most SMBH mergers in FLARES fall within the mass range $10^{5}~{\rm M}_{\sun} \la M_{\rm BH} \la 10^{8}~{\rm M}_{\sun}$, which is within LISA's target range, we investigate their detectability in future LISA observations. In addition, we examine how unresolved physical processes in current cosmological simulations -- such as sub-kpc-scale evolution, SMBH spins, and SMBH seed implementation -- affect these predictions.

To model the GW signal of each SMBH merger, we use the SMBH masses and merger redshifts from FLARES as inputs and compute the signals using the Binary Observability With Illustrative Exploration (\textsc{bowie}) python package\footnote{\href{https://github.com/mikekatz04/BOWIE}{https://github.com/mikekatz04/BOWIE}} \citep{Katz2019}. \textsc{bowie} incorporates phenomenological waveform models for non-precessing (aligned-spin\footnote{That is, the PhenomD model is restricted to systems where the black hole spins are parallel or anti-parallel to the direction of the orbital angular momentum. For SMBH spin orientations in binaries, it has been suggested that in gas-rich galactic environments -- more common at high redshifts -- the orbital and spin orientations of both SMBHs tend to align with the large-scale gas flow, while in gas-poor environments, the spin orientations are more likely to be isotropic \citep[see e.g.][]{Bogdanovic2007,Miller2013,Gerosa2015,Steinle2023,Bourne2024}.}) black hole binaries, specifically the PhenomD model \citep{Husa2016,Khan2016}, which provides the GW signal across the inspiral, merger, and ringdown phases. The PhenomD waveform model has been calibrated using numerical relativity simulations for mass ratios down to $q = 1:18$, with mismatch errors typically less than 1 per cent in the calibration region, and it produces physically reasonable results beyond the calibration region; see section IX of \citet{Khan2016} for detailed discussions.

To assess the detectability of each SMBH merger by LISA, we use the GW signal (i.e. characteristic strain, which measures the strength of a GW signal, as a function of observed frequency) returned from PhenomD, $h_{\rm c}(f)$, and adopt the proposed LISA sensitivity curve \citep{Amaro-Seoane2017} \citep[i.e. the `PL' sensitivity curve from][]{Katz2019}, $h_{\rm PL}(f)$. The signal-to-noise ratio (SNR) is estimated by integrating the ratio between $h_{\rm c}$ and $h_{\rm N}$ over the frequency space \citep{Katz2019},
\begin{equation}
    {\rm SNR} = \left[2 \times \frac{16}{5} \int_{f_{\rm start}}^{f_{\rm end}} \frac{1}{f} \frac{h_{\rm c}^2(f)}{h_{\rm N}^2(f)} {\rm d}f \right]^{1/2},
\end{equation}
where the factor of $2$ arises from LISA's design as a two-channel detector, the factor of $16/5$ originates from averaging over all possible source orientations and sky locations, and $f_{\rm start} = f(t_{\rm start})$ and $f_{\rm end} = f(t_{\rm end})$ correspond to the GW frequencies at the beginning and end of the LISA observation window. Following previous works \citep[e.g.][]{Banks2022,Chen2022}, we adopt the most optimistic estimate of the SNR for each SMBH merger by taking $t_{\rm end}$ as the end of the merger phase and setting $t_{\rm start}$ to $T_{\rm obs}$ before $t_{\rm end}$, thereby integrating the segment of the waveform with the largest strains. In reality, the SNR could be lower if the LISA observation window does not fully overlap with the merger period of the binary \citep[see e.g.][for more discussions]{Katz2020}. In this study, we assume an observation time of $T_{\rm obs} = 4$ yr for LISA. The noise term, $h_{\rm N}$, in the denominator of the integrand is the maximum of the LISA sensitivity curve and the background noise of unresolved Galactic white dwarfs \citep{Bender1997,Hiscock2000} at each frequency, i.e. $h_{\rm N}(f) = \max [h_{\rm PL}(f), h_{\rm GB}(f)]$ (see Figure~\ref{fig:gw_dep} for more details).

Note that the default cosmological parameters used in \textsc{bowie} are based on the Planck15 set \citep{Planck2016}. For all calculations presented below, we have replaced these with the Planck13 set \citep{Planck2014} to maintain consistency with FLARES, although this change has a negligible effect. Unless otherwise specified, we have assumed zero spins for SMBHs in our calculations, as FLARES does not model SMBH spin evolution. In this study, the dimensionless spin parameter is defined as
\begin{equation}
    \chi = \frac{c}{G}\frac{\mathbfit{S} \cdot \hat{\mathbfit{L}}}{M_{\rm BH}^2},
\end{equation}
where $\mathbfit{S}$ is the SMBH spin vector and $\hat{\mathbfit{L}}$ denotes the direction of the orbital angular momentum. Positive and negative values of $\chi$ indicate that the SMBH spin is aligned and anti-aligned, respectively, with the orbital angular momentum.

In Figure~\ref{fig:gw_dep}, we illustrate the dependence of GW signals on the properties of SMBH mergers, covering the typical parameter space in FLARES. Panel (a) shows characteristic strains for SMBH mergers with different total masses, ranging from $10^{4}~{\rm M}_{\sun}$ to $10^{9}~{\rm M}_{\sun}$, assuming $q=1$, $z_{\rm merger} = 5$, and zero spins. The SNR values, calculated including the Galactic white dwarf background noise \citep{Bender1997,Hiscock2000}, are also provided. More massive SMBH mergers produce stronger GWs at lower frequencies. SMBH mergers with $M_{\rm tot} \la 10^{8}~{\rm M}_{\sun}$ can be detected by LISA with high SNRs ($\geq 10$), whereas more massive mergers ($M_{\rm tot} \ga 10^{9}~{\rm M}_{\sun}$) completely fall outside the detection range of LISA. For the SMBH merger with $M_{\rm tot} = 10^{4}~{\rm M}_{\sun}$, only the inspiral phase is observable by LISA, while for the merger with $M_{\rm tot} = 10^{8}~{\rm M}_{\sun}$, only parts of the merger and ringdown phases can be detected. Panel (b) shows that SMBH mergers with higher mass ratios ($q \sim 1$) emit stronger GWs. For non-spinning SMBHs merged at $z_{\rm merger} = 5$ with $M_{\rm tot} = 10^{6}~{\rm M}_{\sun}$, only those with mass ratios $q \ga 10^{-2}$ are observable by LISA. The merger with $q = 10^{-3}$ has a GW signal comparable to the Galactic background, while the one with $q = 10^{-4}$ is even weaker than the Galactic background noise. Compared to $M_{\rm tot}$ and $q$, the parameters $z_{\rm merger}$ in the plotted range and $\chi$ have relatively smaller effects on GW signals, as illustrated in Panels (c) and (d). Panel (c) demonstrates that SMBH mergers occurring at earlier times have more redshifted GW frequencies and lower characteristic strains. In Panel (d) and in this study, we focus primarily on equal-spin cases, as the PhenomD model was calibrated using mostly equal-spin waveforms and its accuracy may be reduced for systems with unequal spins \citep{Khan2016}.

\subsection{LISA detectability and environmental dependence}\label{subsec:lisa_det}

\begin{figure} 
\centering\includegraphics[width=\columnwidth]{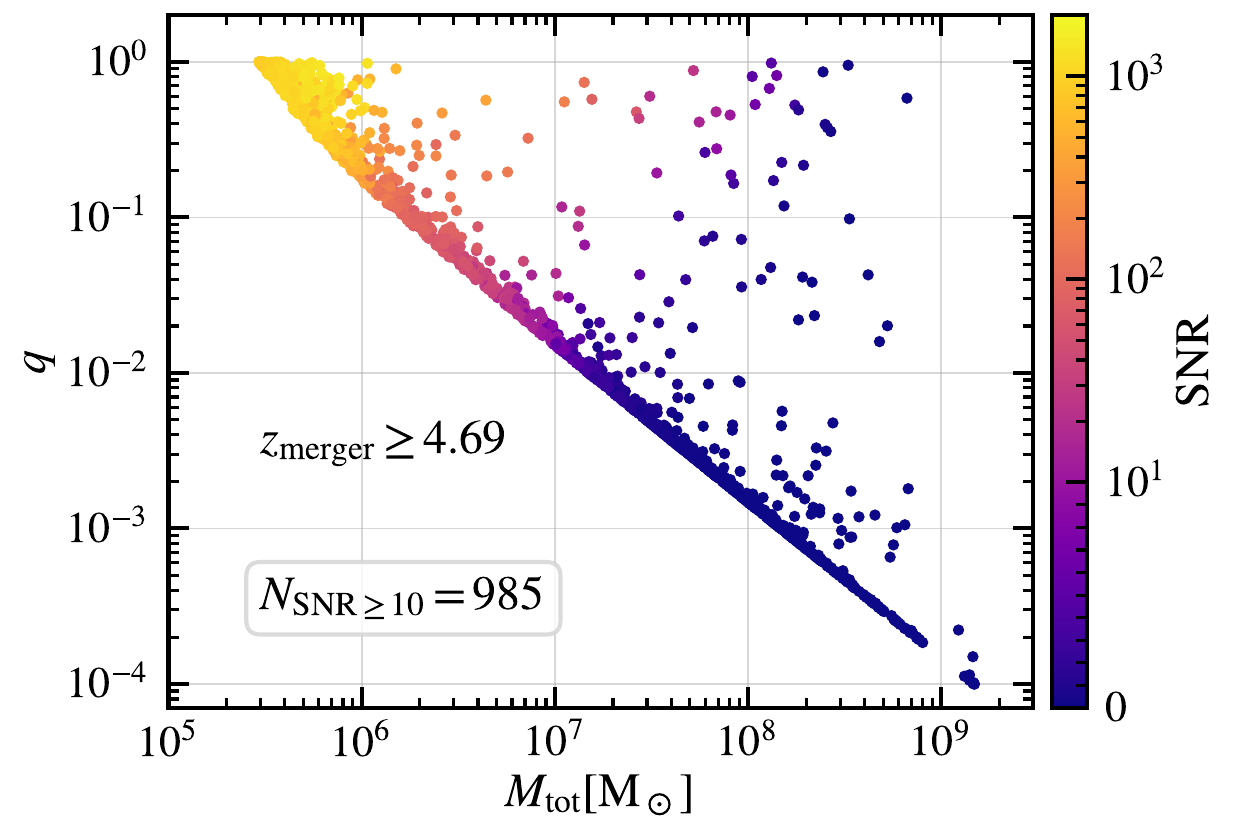}
\caption{Predicted SNR of all FLARES high-redshift SMBH mergers in LISA observations. The 2,017 mergers are plotted on the $q$--$M_{\rm tot}$ plane, with colours indicating SNR. Note that the colour bar uses a linear scale for ${\rm SNR} < 10$ and a logarithmic scale for ${\rm SNR} \geq 10$. In total, 985 mergers have ${\rm SNR} \geq 10$. These observable mergers tend to have lower total masses and higher mass ratios.}
\label{fig:snr_Mtot_q}
\end{figure}

\begin{figure} 
\centering\includegraphics[width=\columnwidth]{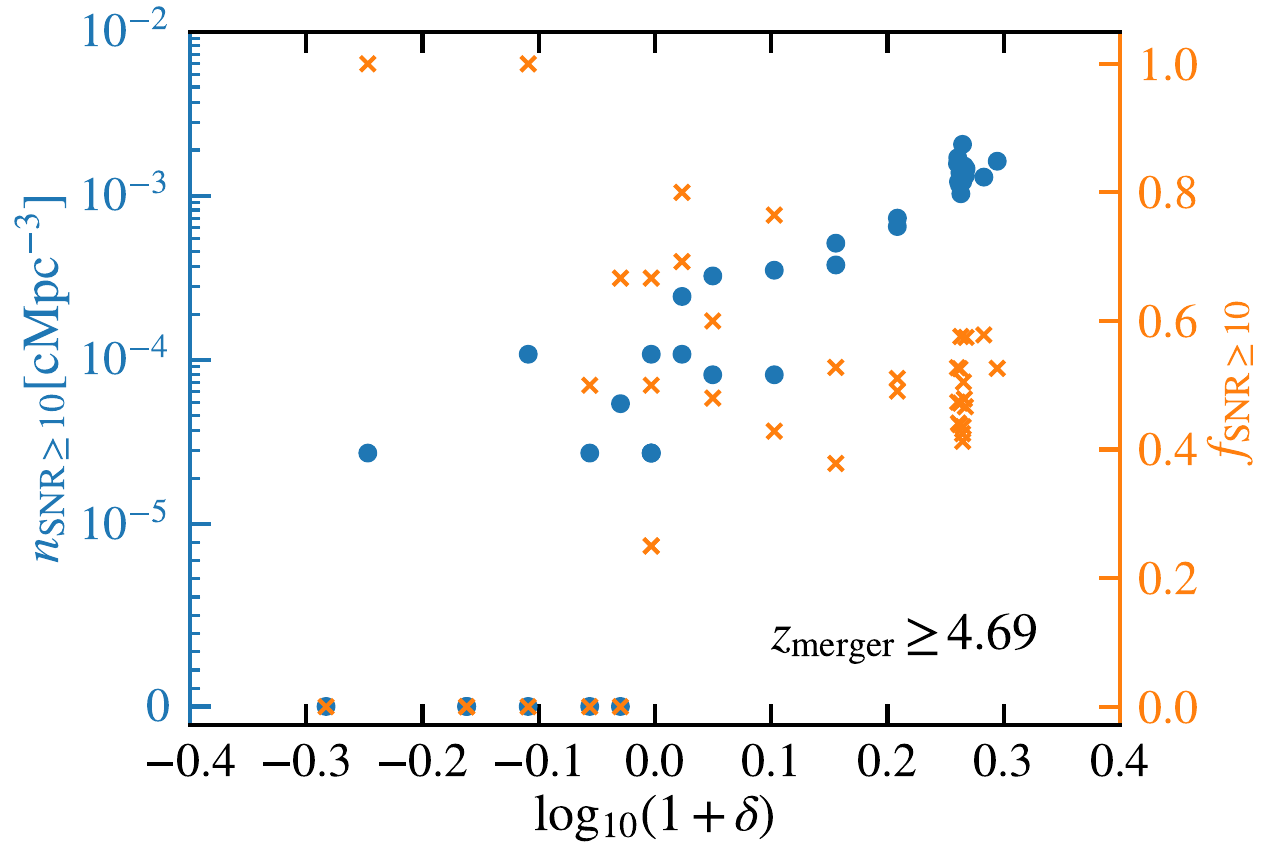}
\caption{Dependence of the number of detectable SMBH mergers on environmental density. Blue dots represent the comoving number density of all detectable SMBH mergers (i.e. those with ${\rm SNR} \geq 10$) across all 40 FLARES regions (left $y$-axis) as a function of environmental overdensity. Orange crosses indicate the fraction of detectable SMBH mergers, $f_{{\rm SNR} \geq 10} = n_{{\rm SNR} \geq 10} / n_{\rm merger}$ (right $y$-axis), also as a function of environmental overdensity. Higher-density regions contain more detectable SMBH mergers, but their fractions of observable SMBHs decrease slightly, as some high-mass or low-mass-ratio SMBH mergers fall beyond the LISA sensitivity range.}
\label{fig:n_lisa_env}
\end{figure}

\begin{figure} 
\centering\includegraphics[width=\columnwidth]{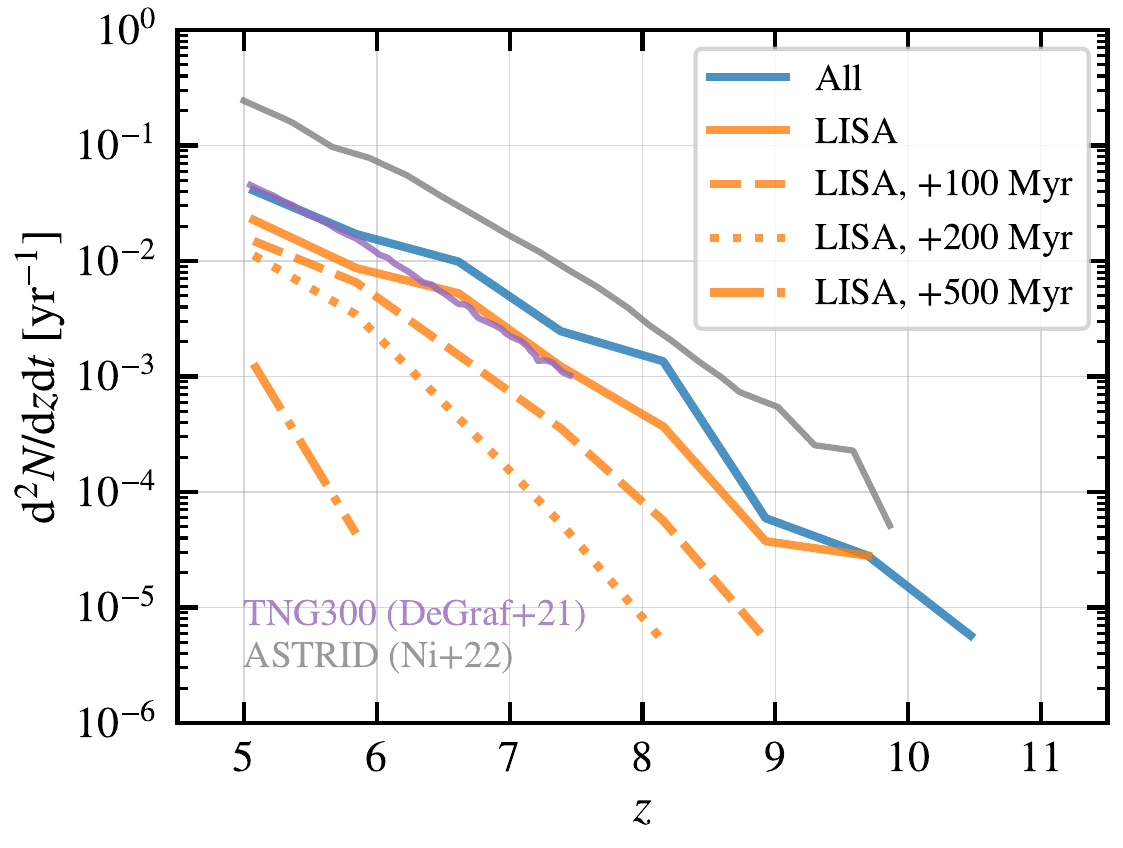}
\caption{SMBH merger rates and predicted LISA detection rates. The blue line represents the merger rates computed from all FLARES SMBH mergers, compare with the results from the TNG300 \citep[purple, ][]{DeGraf2021} and ASTRID \citep[gray,][]{Ni2022} simulations. The orange solid curve shows the predicted detection rate for LISA (i.e. computed from those SMBH mergers with ${\rm SNR} \geq 10$) as a function of merger redshift. The orange dashed, dotted, and dash-dotted lines indicate the predicted LISA detection rates after incorporating constant merger time delays of 100, 200, and 500 Myr, respectively. The unresolved SMBH merger time-scale in current cosmological hydrodynamical simulations significantly affects the predicted GW detection rate, particularly at high redshifts.}
\label{fig:det_rate}
\end{figure}

The SNRs of all 2,017 SMBH mergers from FLARES are plotted as a function of $M_{\rm tot}$ and $q$ in Figure~\ref{fig:snr_Mtot_q}. SMBH mergers with lower total masses and higher mass ratios exhibit higher SNRs, with the highest SNRs (${\sim} 1000$) occurring in mergers where both SMBH masses are close to the seed mass. This highlights the importance of SMBH seeding models in predicting the GW detections, which we discuss in more detail in Section~\ref{subsec:dis}. Assuming a detection threshold of ${\rm SNR} \geq 10$, a total of 985 SMBH mergers in FLARES are detectable by LISA.

In Figure~\ref{fig:n_lisa_env}, we further examine the dependence of detectability on environments. The blue dots show the comoving number density of detectable SMBH mergers ($n_{{\rm SNR} \geq 10} = N_{{\rm SNR} \geq 10} / V_{\rm c}$) as a function of overdensity. Here, $N_{\rm SNR \geq 10}$ denotes the number of SMBH mergers with ${\rm SNR} \geq 10$ in a FLARES region. As expected, denser regions exhibit a higher number density of observable events. The orange crosses represent the fraction of SMBH mergers that are observable ($f_{\rm SNR \geq 10} = N_{\rm SNR \geq 10} / N_{\rm merger}$) for all FLARES regions. There is a weak trend where higher-density regions have lower fractions of SMBH mergers that are observable ($f_{\rm SNR \geq 10} \sim 0.5$). This is because some high-mass or low-mass-ratio SMBH mergers, which are more common in dense regions, fall beyond LISA’s sensitivity range, as demonstrated above.

Once we identify which SMBH mergers in a FLARES region are detectable by LISA, we can compute the detection rates by incorporating the weight of each region. The GW detection rate is defined as
\begin{equation}\label{eq:detection_rate}
    \frac{{\rm d}^2 N}{{\rm d}z{\rm d}t} = \frac{1}{\Delta z}\int_{z}^{z + \Delta z} \frac{{\rm d}^{2} N(z)}{{\rm d}z{\rm d}V_{\rm c}} \frac{{\rm d}z}{{\rm d}t} \frac{{\rm d}V_{\rm c}}{{\rm d}z}\frac{{\rm d}z}{1 + z},
\end{equation}
which gives the number of observed SMBH binaries (merged between redshifts $z$ and $z+\Delta z$) per unit redshift per unit time. The first term in the integrand can be estimated from simulations as
\begin{equation}
    \frac{{\rm d}^{2} N(z)}{{\rm d}z{\rm d}V_{\rm c}} = \sum_i w_i \frac{N_{i}(z, z+\Delta z)}{\Delta z V_{{\rm c}, i}},
\end{equation}
where $N_{i}(z, z+\Delta z)$ is the number of detectable SMBH mergers in the $i$-th FLARES region occurring in the redshift bin $[z, z+\Delta z)$, $V_{{\rm c}, i}$ is the comoving volume of the FLARES spherical region, and $w_i$ is the weight assigned to this region when computing composite quantities. The other terms in the integrand of Equation~(\ref{eq:detection_rate}), ${\rm d}z / {\rm d} t$ and ${\rm d}V_{\rm c} / {\rm d} z$, can be computed analytically, and they are functions of cosmological parameters. 

The total SMBH merger rates (i.e. considering all mergers with ${\rm SNR} \geq 0$) and the LISA detection rates (i.e. including only mergers with ${\rm SNR} \geq 10$) as a function of redshift are shown in Figure~\ref{fig:det_rate}, represented by blue and orange solid curves, respectively. Within the redshift range $5 \la z \la 9$, the ratio between the detection and total merger rates is $\sim 0.5$, indicating that roughly half of SMBH mergers can be detected by LISA.

For comparison, we also plot the total SMBH merger rates from the TNG300 \citep{DeGraf2021} and ASTRID \citep{Ni2022} simulations, which, compared to other periodic cosmological hydrodynamical simulations with uniform resolution, have larger box sizes and provide better statistics for galaxies in the early Universe. The SMBH merger rates from FLARES agree well with those from the TNG300 simulation. The TNG300 simulation, with a box size of 302.6 cMpc, is the largest among all TNG simulations \citep{Pillepich2018} but also has the lowest particle resolution, a factor $\sim 10$ lower than FLARES. The earliest SMBH merger in TNG300 occurred at $z = 8.18$. In comparison, the earliest SMBH merger in FLARES took place at $z = 10.85$. As a result, FLARES enables us to compute the SMBH merger rates at even earlier cosmic times. With a box size of 369.1 cMpc, the ASTRID simulation predicts systematically higher SMBH merger rates -- by a factor of a few -- compared to FLARES and TNG across the redshift range shown in Figure~\ref{fig:det_rate}. This discrepancy likely arises primarily from differences in SMBH seeding prescriptions. For example, ASTRID assigns seed masses following a power-law distribution, allowing for lower-mass seeds and resulting in a higher abundance of seed SMBHs in low-mass haloes compared to the fixed-mass seeding adopted in FLARES and TNG. See \citet{Ni2022} for further details on the ASTRID SMBH model and the discussions in \citet{DeGraf2024}.

By integrating the LISA detection rate (orange solid curve in Figure~\ref{fig:det_rate}) over the redshift range $z = 4.69$ to $10.85$ (covered by FLARES SMBH mergers), we estimate that LISA will detect SMBH mergers at $z \geq 4.69$ at a rate of $0.030~{\rm yr}^{-1}$. This suggests that detections of SMBH mergers with $M_{\rm tot} \ga 10^{5}~{\rm M}_{\sun}$ at these early cosmic times will be rare but still within reach for LISA over its planned mission lifetime (i.e. at least 4 years but potentially up to 10 years, according to \citealt{Amaro-Seoane2017,Colpi2024}).

As FLARES utilizes the same galaxy formation subgrid model as EAGLE but provides better statistics for SMBH properties at high redshifts by resimulating certain overdense regions, the LISA detection rate estimates from FLARES can be regarded as complementary to those from EAGLE \citep{Salcido2016} at $z \ga 5$. It is important to note that the FLARES estimate does not represent the total detection rate for the LISA observatory, as SMBH mergers at lower redshifts contribute additional -- and likely dominant -- detections \citep[see e.g.][]{Salcido2016,Katz2020,Volonteri2020,Chen2022,Li2022,Wang2025}. Compared to previous cosmological simulations that evolve to lower redshifts, the FLARES predicted detection rate is consistently lower. For example, \citet{Salcido2016} predict an eLISA detection rate of ${\sim} 2~{\rm yr}^{-1}$ after considering all SMBH mergers down to $z=0$ in EAGLE; \citet{Katz2020} report a rate of ${\sim}0.5$--$1~{\rm yr}^{-1}$ using SMBH mergers to $z=0$ from Illustris; The ASTRID simulations yield predicted detection rates of $0.3$--$0.7~{\rm yr}^{-1}$ for $z>3$ SMBH mergers \citep{Chen2022}, and $5.6$--$10.5~{\rm yr}^{-1}$ for $z \geq 0$ mergers \citep{Wang2025}.

We further note that in FLARES, due to the seeding prescription, only SMBHs with $M_{\rm BH} \geq 10^{5}~h^{-1}{\rm M}_{\sun}$ are considered, and thus mergers involving low-mass SMBHs (i.e. $10^{3} \la M_{\rm BH}/{\rm M}_{\sun} \la 10^{5}$) are absent in the simulations. Recent studies have shown that mergers of low-mass SMBHs could make a significant contribution to LISA detections \citep[see e.g.][]{Chen2022,Bhowmick2024b,DeGraf2024,McCaffrey2025}. Therefore, the FLARES prediction here should be interpreted as a lower limit on the LISA detection rate for SMBH mergers at high redshift.

\subsection{Discussions} \label{subsec:dis}

In this subsection, we discuss the impact of unresolved physical processes in cosmological simulations that could affect GW observations. These include the coalescence of SMBH binaries at separations below the resolution limit (${\sim}1$ kpc), SMBH seed formation, and SMBH spin evolution.

\subsubsection{Impact of merger time delays}

The coalescence of two SMBHs typically proceeds through three phases: (i) During a galaxy collision, at ${\sim}$ kpc scales, two SMBHs lose their orbital energy and angular momentum due to dynamical friction from dark matter, stars, and gas \citep{Chandrasekhar1943,Ostriker1999}, eventually forming a gravitationally bound SMBH binary. (ii) At ${\sim}$ pc scales, the SMBH binary continues to shrink via interactions with stars through the gravitational slingshot mechanism \citep[three-body interactions; see e.g.][]{Quinlan1996} and with a circumbinary gas disc \citep[see][for a review]{Lai2023}. (iii) Once the separation reaches ${\sim}$ mpc scales, GW emission becomes the dominant mechanism for dissipating orbital energy and angular momentum, leading to the final coalescence \citep{Peters1963,Peters1964}.

As detailed in Section~\ref{subsec:subgrid}, in FLARES, two SMBHs are merged instantaneously when their separation is less than both the SMBH smoothing length ($h_{\rm BH}$) and three gravitational softening lengths, provided their relative velocity is smaller than the circular velocity at the radius of $h_{\rm BH}$. Before this ad hoc merger, the SMBH separation is typically at ${\sim}$ kpc or sub-kpc scales. Thus, like other state-of-the-art cosmological simulations, FLARES only resolves the first phase of coalescence, while the second and third phases remain unresolved. Consequently, the actual SMBH mergers likely occur later than the redshifts recorded in FLARES.

The time-scale following the first phase remains one of the largest uncertainties in SMBH merger studies \citep[see e.g.][]{Amaro-Seoane2023}. Direct N-body simulations of gas-free galaxy collisions, which resolve non-softened gravity and three-body interactions, suggest that SMBH mergers typically occur on ${\sim}$ Gyr time-scales \citep[e.g.][]{Khan2012,Vasiliev2015}. For high-redshift galaxies, however, gas plays a crucial role. At $z \sim 5$, the age of the Universe is only ${\sim} 1$ Gyr, and it is essential for other processes to facilitate high-redshift SMBH coalescence. Recent RABBITS simulations by \citet{Liao2024a,Liao2024b}, which resolve non-softened gravity while incorporating hydrodynamics and other galaxy formation processes (e.g., radiative cooling, star formation, stellar feedback, and AGN feedback), introduce a circumbinary disc accretion subgrid model \citep{Liao2023}. These simulations show that in gas-rich galaxy mergers, the SMBH coalescence time-scales after the first phase are typically ${\sim} \, 10^2$ Myr, due to nuclear star formation that efficiently refills the loss cone and facilitates the merging process \citep{Liao2024a}. However, the exact coalescence time-scale depends on galaxy properties and the subgrid modeling of galaxy formation processes. See also, e.g. \citet{Li2024,Zhou2025} for recent progress in this direction.

To address the impact of merger time delays, previous works based on modern cosmological galaxy formation simulations have adopted post-processing models \citep[e.g.][]{Kelley2017,Katz2020,Volonteri2020,Banks2022,Chen2022,Li2022}. These studies typically employ semi-analytical modeling of the time-scales or orbital evolution across different SMBH binary evolutionary phases -- including dynamical friction, stellar hardening, viscous torques from circumbinary discs, and gravitational wave emission -- coupled with host galaxy properties, to provide more realistic estimates of merger delays. The SMBH merger time-scales reported in these works range from ${\sim} 10^{-2}$ to ${\sim} 10^{3}$ Gyr, depending on the detailed SMBH/galaxy properties and the adopted delay time models. In our analysis, however, the limited time resolution of FLARES snapshots hinders accurate crossmatching of host galaxies and estimation of their properties at the time of the merger. Therefore, we adopt a constant delay time model to address the impact of merger delays, following the approach of \citet{Salcido2016,DeGraf2021,Bhowmick2024c}.

Specifically, we apply three constant time delays -- $100$ Myr, $200$ Myr, and $500$ Myr -- to FLARES SMBH mergers and plot the resulting LISA detection rates using dashed, dotted, and dash-dotted lines, respectively, in Figure~\ref{fig:det_rate}. We find that merger time delays significantly affect GW detections, particularly at high redshifts. With time delays of $100$ Myr, $200$ Myr, and $500$ Myr, the integrated LISA detection rates for $z \geq 4.69$ reduce to $0.018~{\rm yr}^{-1}$, $0.012~{\rm yr}^{-1}$ and $0.001~{\rm yr}^{-1}$, respectively. The cosmic time between the first SMBH merger in FLARES ($z = 10.85$) and the end redshift of simulations ($z = 4.69$) is ${\sim}850$ Myr. Therefore, if the merger time delay exceeds 850 Myr, none of the FLARES SMBH `binaries' will coalesce within the simulation redshift range.

Therefore, improving the modeling of merger time-scales is crucial for cosmological simulations to provide more accurate LISA detection rate predictions for high-redshift SMBH mergers \citep[see also e.g.][]{Volonteri2020,DeGraf2021,Banks2022,Chen2022,Bhowmick2024c}. Of course, the constant delay model used here is an oversimplification, and the results should be regarded as illustrative estimates rather than quantitatively robust predictions. We further note that as discussed in Section~\ref{subsec:mass_ratio}, the lack of accurate SMBH dynamical evolution can also affect the distribution of merging SMBH masses and thus the resulting GW predictions. This limitation may also influence the dependence of detectable SMBH mergers on environment (Figure~\ref{fig:n_lisa_env}), especially in high-density regions where galaxy interactions are more complex and frequent. In our future work, we plan to incorporate the numerical methods from the RABBITS simulations into FLARES to improve the modeling of SMBH coalescence and to increase the time resolution of snapshot outputs, thereby enabling a more detailed study of the relationship between SMBH mergers and their host galaxies.

\subsubsection{Impact of SMBH seed models}

The formation of SMBH seeds remains an open question in astrophysics \citep[see e.g.][for a review]{Volonteri2021}. Several competing (or possibly complementary) formation mechanisms have been proposed, including remnants of Population III stars, runaway mergers in dense star clusters, and direct collapse of Lyman-Werner irradiated gas.

Similar to many other cosmological simulations, and due to the limitation in numerical resolution and the uncertainties in SMBH seed formation, FLARES adopts a simplified subgrid model: an SMBH seed with mass $m_{\rm seed} = 10^{5} ~ h^{-1}{\rm M}_{\sun}$ is added at the centre of an FOF halo when its mass reaches $10^{10} ~ h^{-1}{\rm M}_{\sun}$, provided it does not already host an SMBH. To investigate the impact of SMBH seeding implementations on GW observations, we use the Region 08 test run (described in Section~\ref{subsec:bh_seed_prop}), which adopts a lighter seed mass of $m_{\rm seed} = 10^{4} ~ h^{-1}{\rm M}_{\sun}$, and compare its SMBH merger SNRs to those from the fiducial run.

\begin{figure} 
\centering\includegraphics[width=\columnwidth]{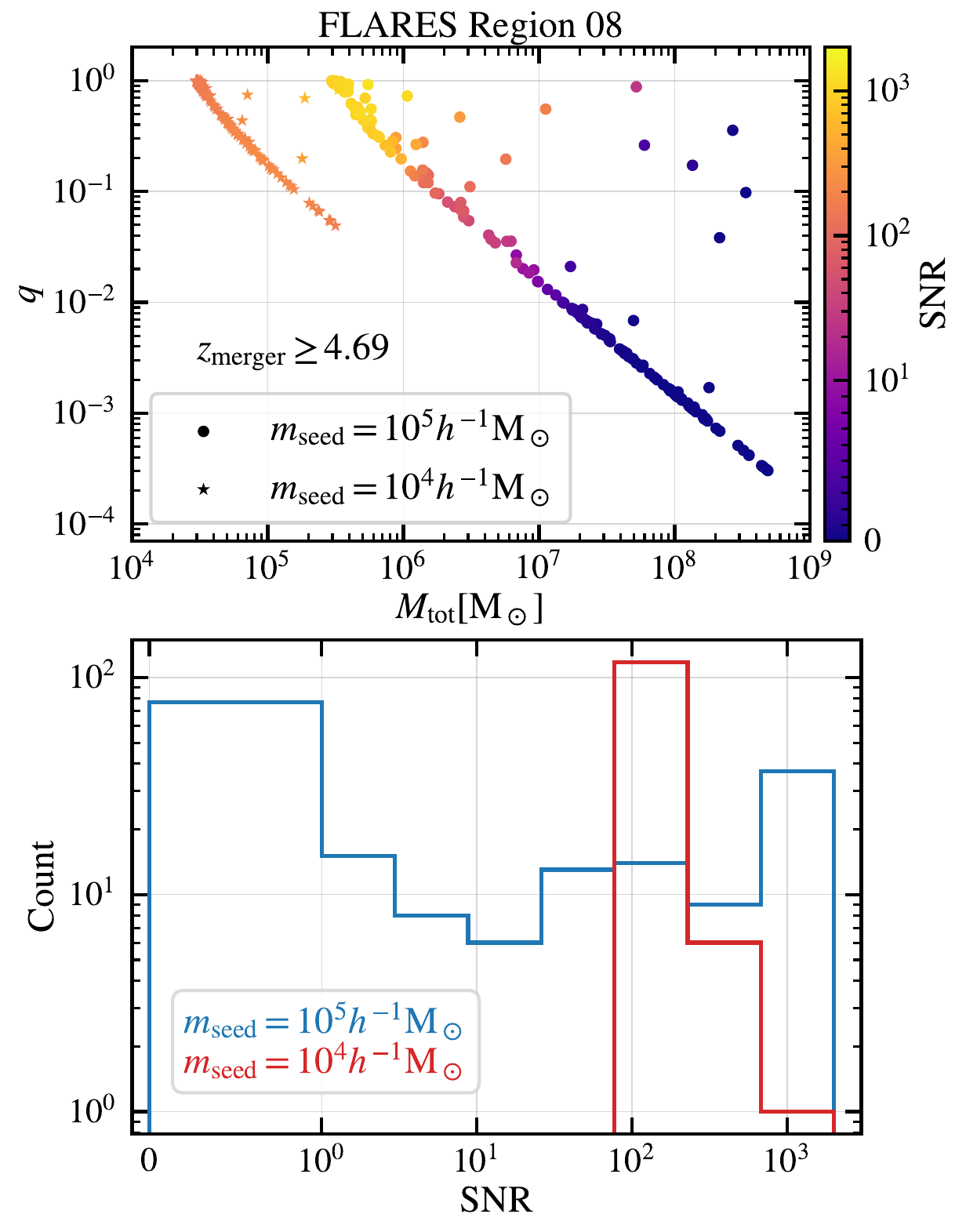}
\caption{Impact of SMBH seed mass on the detectability of GW signals. {\it Top}: Predicted SNR for LISA detections as a function of $M_{\rm tot}$ and $q$ from FLARES Region 08 runs with two different SMBH seed masses, i.e. $m_{\rm seed} = 10^{5} ~ h^{-1} {\rm M}_{\sun}$ (circles) and $m_{\rm seed} = 10^{4} ~ h^{-1} {\rm M}_{\sun}$ (stars). {\it Bottom}: Histograms of SNR. Blue and red represent the higher and lower SMBH seed masses, respectively. With a lower SMBH seed mass, simulated SMBH mergers tend to have lower total masses and higher mass ratios, making all mergers observable by LISA (i.e. ${\rm SNR} \geq 10$).}
\label{fig:snr_seed_mass}
\end{figure}

In Figure~\ref{fig:snr_seed_mass}, we present the SNRs of SMBH mergers from these two runs. The top panel shows the results in the $q$--$M_{\rm tot}$ plane, while the bottom panel provides the distributions of SNRs. As discussed in Section~\ref{subsec:bh_seed_prop}, lighter seeds lead to lower SMBH masses and higher mass ratios. Consequently, all 124 SMBH mergers in the test run have ${\rm SNR} \geq 10$, making them detectable by LISA. This contrasts with the fiducial run, where only 76 out of 179 SMBHs ($42.5 \%$) are detectable. We also note that the fiducial run contains more mergers with very high SNR ($\ga 10^3$). These mergers typically have $M_{\rm tot} \sim 3 \times 10^{5}~{\rm M}_{\sun}$ and $q \sim 1$, i.e. nearly equal-mass mergers between SMBHs close to the seed mass. This suggests that the fiducial seed mass coincides with LISA's peak sensitivity at high redshifts.

From this comparison, we can conclude that the SMBH seed mass has a significant impact on the predictions of high-redshift GW events detectable by LISA, echoing the conclusions of e.g. \citet{Bhowmick2024b} and \citet{DeGraf2024}. Note that, in this analysis, we only varied the SMBH seed mass, representing one of the simplest SMBH seeding tests. More sophisticated seeding models incorporate additional criteria -- such as gas density, metallicity, and galactic environment -- which better reflect the physical conditions involved in SMBH seed formation \citep[e.g.][]{Taylor2014,Tremmel2017,Wang2019,DeGraf2020,Bhowmick2022,Bhowmick2024a,Bhowmick2024b,Bhowmick2024c,McCaffrey2025}. These additional criteria can have additional or even more significant impact on SMBH merger and GW detection rates \citep[e.g.][]{DeGraf2020,Bhowmick2024b, Bhowmick2024c}. For example, \citet{Bhowmick2024c} demonstrate that considering more realistic conditions for direct collapse seed formation -- such as requirements for Lyman-Werner flux, low gas spin, and high-density halo environments -- can cause SMBH merger rates to vary by nearly two orders of magnitudes at $z > 5$. Together, these findings imply that future LISA observations will place stringent constraints on the SMBH seeding models adopted in cosmological simulations.

\subsubsection{Impact of SMBH spins}

According to the no-hair theorem of general relativity \citep{Israel1967,Israel1968,Carter1971,Hawking1972,Robinson1975}, spin is one of the three fundamental properties of a black hole, with the other two being mass and charge. Astrophysical black holes are usually assumed to be uncharged, as they can be effectively neutralized by discharge to their surrounding environment. Thus, they are characterized only by mass and spin \citep[see e.g.][]{Abramowicz2013}. While SMBH mass growth is modeled in FLARES, the spin evolution is completely neglected, as is common in many cosmological simulations. 

Recent observations using the X-ray reflection spectroscopy method suggest that most SMBHs with $M_{\rm BH} < 3 \times 10^{7}~{\rm M}_{\sun}$ tend to be rapidly spinning ($|\chi| > 0.9$), whereas some modestly spinning SMBHs appear at higher masses \citep[see][for a review]{Reynolds2021}. But note that these results may be affected by selection bias and might not fully represent the SMBH population in the Universe. As observed in Figure~\ref{fig:gw_dep}, SMBH spins have a relatively weaker effect on GW signals compared to SMBH mass. To quantitatively assess the impact of spins on our predictions of LISA detection rates, we consider two representative cases, $\chi_1 = \chi_2 = 1$ and $\chi_1 = \chi_2 = -1$, motivated by the aforementioned observational results. Following the case of $\chi_1 = \chi_2 = 0$ in Section~\ref{subsec:lisa_det}, we compute the LISA detection rates for SMBH mergers at $z \geq 4.69$, again assuming the detection SNR threshold of ${\rm SNR} \geq 10$. The resulting integrated detection rates are $0.044~{\rm yr}^{-1}$ for aligned spins and $0.027~{\rm yr}^{-1}$ for anti-aligned spins. Compared to the non-spinning case discussed in Section~\ref{subsec:lisa_det} ($0.030~{\rm yr}^{-1}$), the detection rate increases by ${\sim}15 \%$ for maximally aligned spins and decreases by ${\sim}10 \%$ for maximally anti-aligned spins.

Note that SMBH spins are also crucial for estimating the GW-induced recoil velocity, which can impact SMBH evolution and galaxy properties \citep[e.g.][]{Campanelli2007,Blecha2008,Blecha2011,Nasim2021,Mannerkoski2022,Rawlings2025}. To improve GW predictions in future simulations, it is essential to implement a subgrid model for SMBH spin evolution \citep[see e.g.][]{Dubois2014,Fiacconi2018,Bustamante2019,Cenci2021,Talbot2021,Husko2022,Koudmani2024,Sala2024}.

\section{Conclusion} \label{sec:summary}

In this work, we utilized FLARES to study the properties of SMBH mergers in the early Universe ($z \ga 5$), their dependence on environment, and the implications for future LISA detections. The extensive coverage of high-overdensity regions in FLARES enables us to obtain a large sample of SMBH mergers at high redshifts, while the region weighting scheme allows us to estimate the cosmic-average statistical properties. In particular, the resimulation of regions spanning a wide range of overdensities ($-0.5 \la \delta \la 1$) in FLARES offers a unique opportunity to study the environmental dependence of high-$z$ SMBH mergers.

The key properties of FLARES SMBH mergers are summarized as follows:

\begin{description}
    \item {\it SMBH mass}. Primary SMBHs ($M_1$) exhibit a flat distribution from the seed mass ($m_{\rm seed} = 1.48 \times 10^{5}~{\rm M}_{\sun}$) up to $10^{8}~{\rm M}_{\sun}$, followed by a steep decline at higher masses. In contrast, secondary SMBHs ($M_2$) peak near the seed mass and decrease sharply at higher masses.
    \item {\it Mass ratio}. The distribution of mass ratios ($q = M_2/M_1$) spans from ${\sim} \, 10^{-4}$ to $1$, peaking at $q \sim 1$ and remaining relatively flat down to $q \sim 10^{-3}$. The lowest $q$ values arise when the most massive SMBHs merge with seed-mass SMBHs.
    \item {\it Merger redshifts}. The distribution of $z_{\rm merger}$ increases monotonically toward lower redshifts, with more SMBH mergers at later times as a result of the rising number of galaxy mergers.
    \item {\it Environmental dependence}. Denser regions host more SMBH mergers, with the number density ($n_{\rm merger}$) scaling with environmental overdensity ($\delta$) as $n_{\rm merger} = 10^{-3.81} (1 + \delta)^{4.78}$. Within the overdensity range covered by FLARES, the comoving number density in the densest region is ${\sim}500$ times that in the least dense region. As expected, denser regions tend to contain primary and secondary SMBHs with higher masses, exhibit lower mass ratios, and have SMBH mergers that occur earlier.
\end{description}

Based on these high-redshift SMBH mergers, FLARES gives the following predictions for the LISA observatory:

\begin{description}
    \item {\it Detectability}. Within the redshift range explored by FLARES, LISA is expected to be sensitive to SMBH mergers with total masses $10^{5} \la M_{\rm tot}/{\rm M}_{\sun} \la 10^{8}$ and mass ratio $q \ga 10^{-2}$. Assuming a signal-to-noise ratio threshold of ${\rm SNR} \geq 10$, FLARES predicts that LISA will detect high-redshift SMBH mergers ($z \ga 5$) at a rate of $0.030~{\rm yr}^{-1}$.
    \item {\it Environmental dependence}. Denser regions exhibit a higher absolute number density of observable events. However, the fraction of SMBH mergers that are observable is slightly lower in these regions. This is because some high-mass or low-mass-ratio SMBH mergers, which are more common in dense regions, are outside the sensitivity range of LISA.
\end{description}

Furthermore, we study the impact of merger time delays, SMBH seed models, and SMBH spins. Our results suggest that the SMBH seed model has a critical impact on the predictions of SMBH merger properties and GW signatures in the early Universe. In addition, uncertainties in the merger time delays, which are not resolved in current cosmological simulations including FLARES, also significantly affect the predicted LISA detection rates at high redshifts. Therefore, to provide better theoretical predictions for LISA observations, it is critical for future cosmological simulations to improve the modeling of SMBH seeds, merger time-scales, and SMBH spin evolution.

\section*{Acknowledgements}
We thank the anonymous referee for helpful comments. S.L. and Z.J. acknowledge the support by the National Natural Science Foundation of China (NSFC) grant (No. 12473015, 12588202). M.G.A.M. acknowledges the support of a Science and Technology Facilities Council (STFC) studentship. We thank the EAGLE team for their efforts in developing the EAGLE simulation code. This research has made use of NASA’s Astrophysics Data System and arXiv. We gratefully thank the developers of the open-source \textsc{python} packages that were used in the data analysis of this work, including {\sc h5py} \citep{Collette2023}, {\sc Matplotlib} \citep{Hunter2007}, {\sc Numpy} \citep{Harris2020}, {\sc Scipy} \citep{Virtanen2020}, {\sc Pandas} \citep{WesMcKinney2010,pandas2022}, {\sc emcee} \citep{Foreman-Mackey2013}, and {\sc bowie} \citep{Katz2019}.

This work used the DiRAC@Durham facility managed by the Institute for Computational Cosmology on behalf of the STFC DiRAC HPC Facility (www.dirac.ac.uk). The equipment was funded by BEIS capital funding via STFC capital grants ST/K00042X/1, ST/P002293/1, ST/R002371/1, and ST/S002502/1, Durham University and STFC operations grant ST/R000832/1. DiRAC is part of the National e-Infrastructure.

We list here the roles and contributions of the authors according to the Contributor Roles Taxonomy (CRediT).\footnote{\url{https://credit.niso.org/}} {\bf Shihong Liao}: Conceptualization, Formal analysis, Investigation, Methodology, Writing – original draft. {\bf Dimitrios Irodotou}: Conceptualization, Data curation, Investigation, Methodology, Writing – review \& editing. {\bf Maxwell G. A. Maltz, Christopher C. Lovell, Aswin P. Vijayan, William J. Roper}: Data curation, Writing – review \& editing. {\bf Peter A. Thomas}: Conceptualization. {\bf Zhen Jiang, Sophie L. Newman,  Paurush Punyasheel, Louise T. C. Seeyave, Sonja Soininen, Stephen M. Wilkins}: Writing – review \& editing.

\section*{Data Availability}
A portion of the data used to produce this work can be found online: \href{https://flaresimulations.github.io/data.html}{flaresimulations.github.io/data}. Much of the analysis used the raw data produced by the simulation which can be made available upon request.

\bibliographystyle{mnras}
\bibliography{ref} 

\appendix

\section{Fitting method} \label{ap:fitting}

In this appendix, we outline the fitting procedure adopted to determine the power-law relation between the SMBH merger number density, $n_{\rm merger}$, and the environmental overdensity, $1 + \delta$, as shown in Figure~\ref{fig:num_dens_env}. We adopt a Bayesian Markov Chain Monte Carlo (MCMC) approach, assuming a Poisson likelihood for the merger counts,
\begin{equation}
    \mathcal{L}(\alpha, \beta) = \prod^{40}_{i=1} \frac{\lambda_{i}(\alpha, \beta)^{N_{{\rm merger},i}} e^{-\lambda_{i}(\alpha, \beta)}}{N_{{\rm merger},i}!},
\end{equation}
where $i$ loops over all 40 FLARES regions. The expected number of mergers in the $i$th region depends on the local overdensity according to
\begin{equation}
    \lambda_{i}(\alpha, \beta) = n_{{\rm merger},i}(\alpha, \beta)V_{\rm c} = 10^{\alpha}\left(1 + \delta_i\right)^{\beta} V_{\rm c},
\end{equation}
and $\alpha$ and $\beta$ are the free parameters. We assume uniform priors for both parameters over the range $(-10, 10)$. The posterior distribution is sampled using the affine invariant MCMC ensemble sampler \citep{Goodman2010} implemented in the Python package {\sc emcee} \citep{Foreman-Mackey2013}.

\begin{figure} 
\centering\includegraphics[width=\columnwidth]{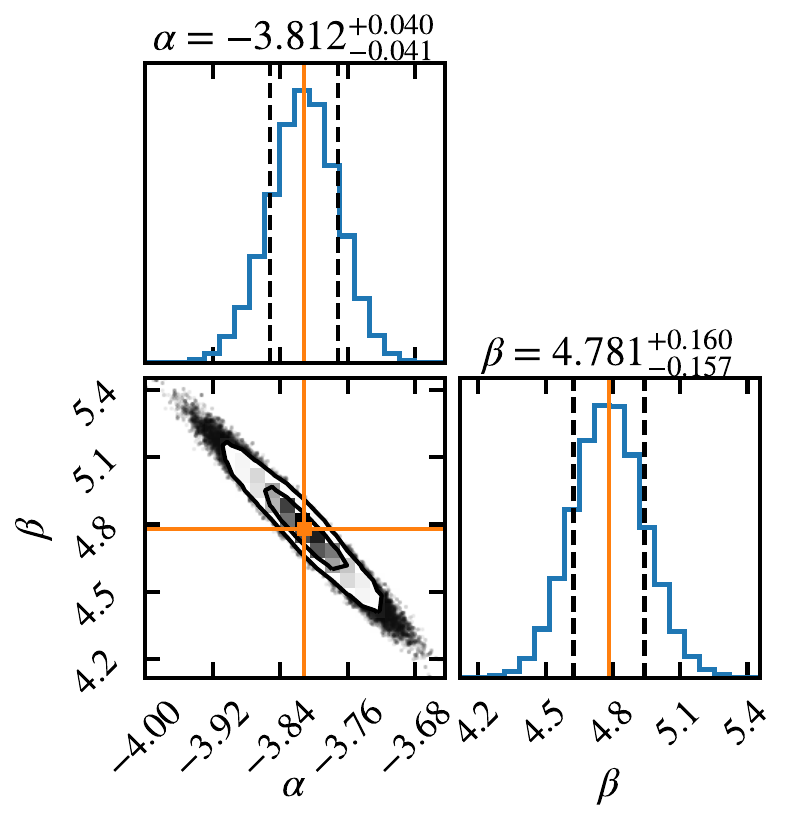}
\caption{Bayesian MCMC fitting results. The bottom-left panel shows the 2D posterior distribution with the $68\%$ and $95\%$ credible contours overplotted. The orange lines indicate the medians of $\alpha$ and $\beta$. The top-left and bottom-right panels show the marginal posterior distributions of $\alpha$ and $\beta$, respectively. The vertical dashed lines mark the 16th and 84th percentiles.}
\label{fig:bayesian}
\end{figure}

The resulting posterior distributions are shown in Figure~\ref{fig:bayesian}. The best-fitting parameters are $\alpha = -3.812^{+0.040}_{-0.041}$ and $\beta = 4.781^{+0.160}_{-0.157}$. The corresponding best-fitting power-law relation is plotted as the blue line in Figure~\ref{fig:num_dens_env}.

\bsp
\label{lastpage}

\end{document}